# A Study of Lightning Activity over Different Ecological Zones of Nepal

**Samin Poudel[a]**

[a] Tribhuvan University, Kathmandu, Nepal, samin.sm@gmail.com

**Abstract**

In the present work, occurrence of lightning activity over different ecological zones of Nepal has been studied. It has been observed that Lower Tropical zone receives lightning strikes with largest value of density of strikes whereas the Trans-Himalayan zone receives lightning strikes with least value of density of strikes. The density of lightning strikes over the Lower Tropical zone is $19.87 \times 10^{-2}$/ km$^2$ per year whereas, that over the Trans-Himalayan is $2.00 \times 10^{-2}$ / km$^2$ per year. Other three zones whose values were observed to be in the higher side were Upper Tropical, Sub-tropical and Water Body with annual densities of lightning strikes of $14.46 \times 10^{-2}$ / km$^2$, $12.05 \times 10^{-2}$ / km$^2$ and $12.00 \times 10^{-2}$ / km$^2$ respectively. During a year, $2.56 \times 10^{-2}$ / km$^2$ and $2.17 \times 10^{-2}$ / km$^2$ are densities of lightning strikes for Alpine and Nival Zones respectively which are close to the lowest value of Trans-Himalayan zone. Remaining two zones, Sub-alpine and Temperate, respectively experienced lightning strikes with densities $6.05 \times 10^{-2}$ / km$^2$ and $8.83 \times 10^{-2}$ / km$^2$ per year.

# Contents

# List of tables

# List of figures

# Acronyms

**GIS**  Geographic Information System

**GLN**  Global Lightning Network

**NAST**  Nepal Academy of Science and Technology

**ESRI**  Environmental Systems Research Institute

**ICIMOD**  International Centre for Integrated Mountain Development

**TOA**  Time Difference of Arrival

**GPS**  Global Positioning System

**CAP**  Central Analysis Processor

**WSI**  Weather Services International

**viii**

# Chapter 1

# Introduction

Lightning, one of the oldest observed powerful natural phenomena on earth, is flash of light created by electric discharge accompanied with tremendous amount of energy. This electrostatic discharge takes place in atmosphere between the electrically charged regions within clouds or between a cloud and the surface of earth to balance the difference between the positive and negative charges. Globally, it strikes the ground about 100 times per second (around 8 million times per year). Lightning, most commonly, occurs within a cloud either from a cloud to the surrounding air or from a cloud to another cloud. Only about 20% of all lightning strikes occur between a cloud and the ground, which is what most people think of as classic lightning. Cloud to ground lightning is the one that is primarily concerned with general people and has been focused in this thesis than other types of lightning.

## 1.1 How does lightning origin?

Different myths, based on various religious beliefs, on the origination of the lightning and thunderstorms have been seen so far. The myths are from different civilizations. Greeks had a belief that the lightning was a weapon of Zeus and the inventor of thunderbolts was Athena, the goddess of wisdom. According to Scandinavian mythology, it was Thor who created lightning with his hammer to attack. Similarly, Hindus had their own belief. They believed that Indra was the god of heaven, lightning, rain, storms and thunder. Indian tribes in North America suggest that it occurred when the mystical thunder bird flaps its wings. Even though these myths are present in written form in various religious books, we cannot find scientific reasons behind them to justify them. From the scientific investigations, conclusions have been drawn that atmospheric convection leading to electric discharge is responsible for lightning (more discussion in 1.2). Lightning occurs when some region of the atmosphere attains an electric charge sufficiently large that the electric field associated with charge cause electrical breakdown of the air [1].

## 1.2 Mechanism and types of lightning

### 1.2.1 Mechanism of charge separation in thunderclouds

The primary source of lightning involving the cloud is cumulonimbus; however, not every cumulonimbus produces lightning discharges. The cumulonimbus which produces lightning discharge is more properly called a thundercloud [2]. As lightning flash is the result of electric discharge and a thundercloud is considered main source for it, thunderclouds are to be electrically charged for this to happen. Different theories have been put forward regarding the charge separation in thundercloud. But, cloud particle collisions are thought to be the main mechanism for cloud electrification [3]. Cloud particle refers to snow, hail, ice crystals, graupel (soft hail) and super cooled water (water below 0 ˚C and above -40 ˚C). In particular, collision between ice crystals and graupel is responsible for the charge separation in thunderclouds. During collision, ice crystals are positively charged where as hail stones are negatively charged. The updrafts during the thunderstorms carry positive crystals to upper region of cloud as they are lighter than hail stones and the heavier wet crystals move toward the base of cloud. Along with the upper and lower charged regions, "In a typical thundercloud a small positive charge is also found below the main negative charge" [4]. Thus developed tri polar structure encourages lightning. Different mechanisms for different types of lightning are briefly discussed in 1.2.2.

### 1.2.2 Types of lightning

Not only natural but also artificially triggered lightning with the purpose of research has come into existence. Here, we focus on the natural forms of lightning. Prospects of different kinds of lightning are being presented in the scientific world recently. But the classification on the basis of the orientation of charge centers, where electric discharge initiates and terminates, is widely accepted. According to this lightning may be categorized into two. One is *cloud to ground discharges* and the other is *cloud discharges.*

#### 1.2.2.1 Cloud to ground discharges

The electric discharge that effectively transfers charge from cloud to the ground is called cloud to ground discharge. From the observed polarity of the charge effectively lowered to ground and the direction of propagation of the initial leader, four different types of lightning discharges between cloud and Earth have been identified [5]. The four types of electric discharge in between cloud and ground are downward negative, upward negative, downward positive and upward positive. As, more than 90 percent of cloud to ground discharge is downward negative, downward negative is simply termed as cloud to ground lightning for its predominance. This form of lightning initiates in a cloud but ends in a cloudless air. During this lightning, negative charges at base of the clouds are transferred towards the positive charge on earth surface from the lower regions of the cumulonimbus cloud. A brief discussion of this form is presented below.

**<u>Downward negative cloud to ground lightning</u>**

Cloud-to-ground (CG) lightning mainly includes processes of preliminary breakdown, stepped leader, first return stroke, inter-stroke process, dart (or dart stepped) leader, subsequent return stroke, and continuous current [6]. The negative charges in the base of thundercloud induce positive charges in the ground which are also essential components for downward negative lightning. To start, the electrons attached to water or ice crystals are discharged from thundercloud and this avalanche of electrons move towards ground by ionising the air in between. The path taken by these electrons to move down is termed as stepped leader, which is 50 meters in length in average. Journey of electrons from cloud to ground is not covered in a single step. Leaders with mean speed of $2 \times 10^5$ m/s [7] and having charge $7 \times 10^{-4}$ C per metre [7] are present. The lengths of the individual steps in stepped leaders vary from 10 to 200m and the inter-step intervals ranges from 40 to 100 μs. Both the step lengths and their brightness increase as the leader speed increase [5].When the stepped leader approaches the ground, an upward leader of positive charges known as upward connecting leader from ground connects to it and thus the stage for the first return stroke is set. The return stroke gives the intensely bright lightning flash and has the greater propagation velocity than that of stepped leader. This return stroke is the actual flow of stroke current that has a median value of about 24000A and is actually the flow of charge from earth to cloud to neutralize the charge centre [8]. Positive charge flowing upward form ground to cloud is equal to the negative that flew down. With pre-channelled path of stepped leader another leader is expected to develop in approximately more than 55 percent [8] of lightning known as dart leader which also comes down with same mechanism as that of stepped. A lightning flash usually have 3-5 strikes with stepped and dart leaders.

Although cloud to ground lightning is less common, it is easier to research and thus best understood. This form of lightning is by far the most damaging and dangerous form of lightning. The sole lightning responsible for the casualties occurring on earth due to lightning is cloud to ground discharge, negative downward lightning in particular.

### 1.2.2.2 Cloud discharges

Cloud discharges occur either within a cloud called intra-cloud lightning or between one cloud to another called inter-cloud lightning. Intra-cloud lightning is the most common type of discharge. Such a discharge takes place in between the two opposite charged regions of the same cloud. As the cloud obscures the lightning flash and makes it hard to see, it has the form of diffuse brightening that flickers. However, on some occasions flash is able to leave the obscuring boundaries of the clouds giving a bright channel of light. Inter-cloud lightning is analogous to intra-cloud lightning. The only difference is that inter-cloud lightning occurs in between the charge centers of two clouds whereas the intra-cloud lightning occurs among the charges of same cloud.

## 1.3 Basic theory

In order to know about electromagnetic field theory involved in lightning phenomenon, first let's go through electrostatic field and magnetostatic field separately in brief.

### 1.3.1 Electrostatic field

The concept of electric charge is the underlying principle for explaining all electrical phenomena [9]. Static electric charges are to be present in and around the region of cumulonimbus cloud for lightning to occur. Static electric charges produce electrostatic field. Although, from old times people have been observing amber attracting tiny pieces of matter in its surrounding under certain conditions, it took some time to understand and explain that this incident was due to electrostatic field. French physicist Charles Auguste de Coulomb, after doing extensive experiments, was able to formulate the interaction forces between electrical charges mathematically. He also made a device to precisely measure these forces. The electrostatic force between two point charges can be written as:

$$\vec{F}_{12} = k \frac{Q_1 \cdot Q_2 \cdot (\vec{r}_2 - \vec{r}_1)}{|\vec{r}_2 - \vec{r}_1|^3} \quad \text{...................... (1.1)}$$

Equation (1.1) can be reported as, "The electrostatic force between two point charges is directly proportional to the amount of electrostatic charge on each of them and inversely proportional to the square of their distance." Here, k is proportionality constant which depends on unit system used and can be obtained from the relation:

$$k = \frac{1}{4\pi \epsilon_0} \quad \text{........................ (1.2)}$$

Where $\epsilon_0 = (8.85 \times 10^{-12})$ AS/V m is the dielectric permittivity of free space. For the force in relation (1.1) to exist presence of electric charges on both geometric objects is must. The forces experienced by charge $Q_1$ which appears due to the electric field of charge $Q_2$ at the position of charge $Q_1$ can mathematically be written as:

$$\vec{E}_2 = \frac{\vec{F}_{21}}{Q_1} = \frac{Q_2 \cdot (\vec{r}_2 - \vec{r}_1)}{4\pi \epsilon_0 |\vec{r}_2 - \vec{r}_1|^3} \quad \text{...................... (1.3)}$$

Moving on to the larger scale, if N point charges are distributed in free space (free space is a linear material in the dielectric sense because the dielectric permittivity of vacuum does not depend on the electric field) the superposition principle can be applied. Thus, the vector sum of all individual fields gives the value of total electric field. Mathematically, total electric field is calculated as:

$$\vec{E}(\vec{r}) = \sum_{i=1}^{N} \frac{Q_1 \cdot (\vec{r} - \vec{r}_i)}{4\pi \epsilon_0 |\vec{r} - \vec{r}_i|^3} \quad \text{...................... (1.4)}$$

### 1.3.2 Magnetostatic field

A magnetic needle which was kept near a current carrying deflected in one side and its direction of deflection was opposite on reversing the direction of current in the wire. This discovery was made by Oersted in 1820. He explained that the magnetic field which sets up around the wire when a current passes through it is responsible for the deflection. Phenomenon of production of magnetic field around a conductor by passing a current through it termed as Oersted's discovery. From Biot-Savart law, we can find the magnetic field $d\vec{B}$ due to a small current carrying element dl as:

$$d\vec{B} = \frac{\mu_0}{4\pi}\frac{Idl\,\sin\theta}{r^2} \quad\quad\quad\quad (1.5)$$

For large numbers of small current carrying elements relation can be written as:

$$\vec{B} = \int d\vec{B} = \frac{\mu_0}{4\pi}\int \frac{Idl\,\sin\theta}{r^2} \quad\quad\quad\quad (1.6)$$

Where,

$\theta$ is the angle between $\vec{dl}$ and $\vec{r}$,

I is the magnitude of current passing through conductor,

($\mu_0 = 4\pi \times 10^{-7}\ WbA^{-1}m^{-1}$), is permeability of vacuum.

Taking divergence of the magnetic field will lead us to one of the Maxwell's equation which is:

$$\triangle \cdot \vec{B} = 0 \quad\quad\quad\quad (1.7)$$

### 1.3.3 Electromagnetic field produced by lightning discharge:

Sections 1.3.1 and 1.3.2 are more than enough to make us believe that both electric and magnetic fields are produced during lightning as static charges and flow of current are the part of this natural phenomenon. Since, different physical processes give rise to different field signatures, the electric and magnetic field signatures so produced are the basic physical parameters in understanding the mechanisms of lightning discharges [2]. Because of the current that flows in the lightning channel in between clouds and ground, an electromagnetic field is radiated from the lightning. Out of the whole electromagnetic energy that is dispersed by the lightning strike, only a minor part is represented by the visible part of spectrum of stroke. Amount of energy being released out during the strike is so huge that the amplitude of the electric field pulse wave is equal to several V/m even at the distance of 100 km. Schematic representation of the lightning channel's assumed geometry is shown in the figure (1.1) in which observation point *P* is the point where the fields are calculated. In order to represent the fields in this geometry, the cylindrical coordinate system is used.

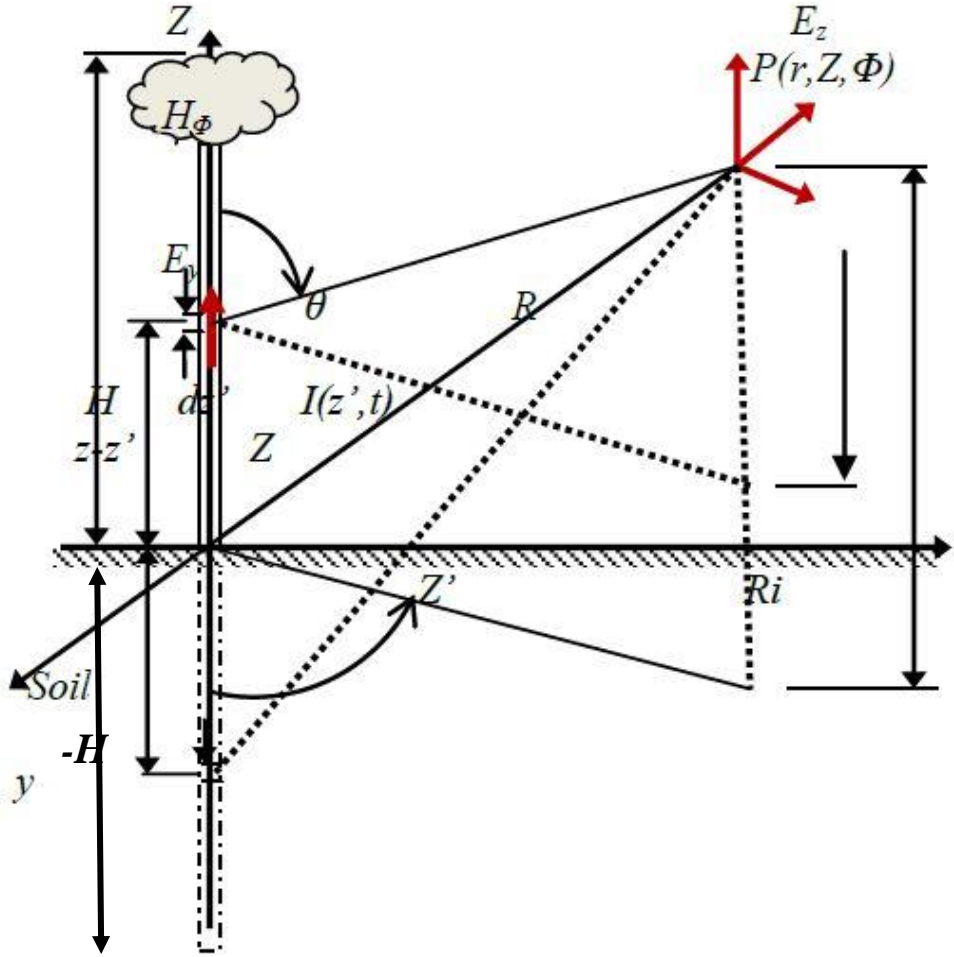

Figure 1.1: Geometrical model used in calculating electromagnetic fields. Adapted from "A New Model of Electromagnetic Fields Radiated by Lightning" [10].

If we consider a perfectly conducting ground, the components of the electric and magnetic fields at the location $P(r, \phi, z)$ produced by a short vertical section of infinitesimal channel $dz'$ at height $z'$ carrying a time-varying current $i(z',t)$ can be computed in the time domain using the following relations [10]:

### 1.3.3.1 Horizontal electric field

The horizontal component of electric field at P ($r$, $\phi$, $z$) can be expressed as:

$$E_r(r,z,t) = \frac{1}{4\pi\epsilon_0}\left[\int_{-H}^{H}\frac{3r(z-z')}{R^5}\int_0^t i(z',\tau - R/c)d\tau dz' + \int_{-H}^{H}\frac{3r(z-z')}{cR^4}i(z',t-R/c)dz' + \int_{-H}^{H}\frac{r^2}{c^2R^3}\frac{\partial i(z',t-R/c)}{\partial t}dz'\right] \quad \text{................ (1.8) [10]}$$

### 1.3.3.2 Vertical electric field

The vertical component of electric field at P ($r$, $\phi$, $z$) can be expressed as:

$$E_z(r,z,t) = \frac{1}{4\pi\epsilon_0}\left[\int_{-H}^{H}\frac{2(z-z')^2-r^2}{R^5}\int_0^t i(z',\tau - R/c)d\tau dz' + \int_{-H}^{H}\frac{2(z-z')^2-r^2}{cR^4}i(z',t-R/c)dz' - \int_{-H}^{H}\frac{r(z-z')^2}{c^2R^3}\frac{\partial i(z',t-R/c)}{\partial t}dz'\right]$$ ............................ (1.9) [10]

### 1.3.3.3 Azimuthal magnetic field

The azimuthal magnetic field at P $(r, \phi, z)$ can be expressed as:

$$B_\phi = \frac{\mu_0}{4\pi}\left[\int_{-H}^{H}\frac{r}{R^3}i(z',t-R/c)dz' + \int_{-H}^{H}\frac{r}{cR^2}\frac{\partial i(z',t-R/c)}{\partial t}dz'\right]$$ ............... (1.10) [10]

Here,

$$R = \sqrt{(z-z')^2 + r^2}$$ ............................. (1.11) [10]

$$H = v\left(t - \frac{R}{c}\right)$$ ............................. (1.12) [10]

In equations 1.8 to 1.12,

*i(z',t)*, is the current carried by the *dz'* dipole at time *t*;

$\varepsilon_0$, is the permittivity of the vacuum;

$\mu_0$, is the permeability of the vacuum;

*c,* is the speed of light;

*R,* is the distance from the dipole to the observation point, and

*r,* is the horizontal distance between the channel and the observation point.

In equations (1.8) and (1.9), the terms containing the integral of the current (charge transferred through *dz'*) are called "electrostatic fields" and, as these terms depend on $\frac{1}{r^3}$ distance, they are the dominant field component near the source whereas terms containing the derivative of the current are called "radiation fields". And, because of their $\frac{1}{r}$ distance dependence, they are the dominant components far from the source. The terms containing the current are called "induction fields". In Equation (1.10), the first term is called "induction magnetostatic field" which is the dominant field component close to the source, and the second term is called "radiation field" which is the dominant field component at far distances from the source. In these equations the presence of the perfectly conducting ground is taken into account by replacing the ground by an equivalent image as shown in figure 1.1. The calculation of the electromagnetic field requires the knowledge of the spatial-temporal distribution of the current along the channel, *i(z',t)* [10].

### 1.4 Effects of lightning

Lightning has accompanied both merits and demerits with it. People have more or less knowledge about the harmful side of the lightning as have heard or faced about its calamities. However, many people are still not clear about the benefits of this powerful force.

Evidence of 250000 - years lightning found in glassy tubes and ancient fulgurites suggest its presence during the time in which the life evolved on earth. Assumptions have been made that lightning was a source to generate the significant molecules like hydrogen cyanide (HCN) which assisted in the evolvement of life in earth. Furthermore, lightning helped to maintain the suitable condition for evolution by strengthening the ozone layer that blocked the harmful radiations like UV radiation. Nitrogen fixation, combination of relatively inert gas with other elements, is essential for the continuation of life on earth and there are not many ways to do it. One of the ways is lightning. Large heat produced during lightning combines atmospheric nitrogen with oxygen to form oxides. These oxides further react with moisture to give nitrates which come on earth with rain.

"It is postulated that the lightning induced fires were mans first source of fire. Fire was of critical importance to humanity: it provided warmth, protection and a means of cooking food" [11]. Lightning ignites the forest fires and plays an important role in the ecological balance. The biologist Mr. Edwin V. Komarek, Sr. in his studies of lightning induced fire damage and the surviving ecology balance indicated that nature's use of lightning fires for clearing dense wooded areas is indeed beneficial to the ecology.

Moving on to the darker side of lightning, it can be considered as a deleterious natural process. Thunderstorms, and lightning in particular, are a major natural hazard to the public, aviation, power companies, and wildfire managers [3]. Every year many fatalities are being reported and the destruction of great deal of property is occurring because of it. The primary cause of death after the lightning strike of a man is cardiopulmonary arrest. Lightning strikes make a man vulnerable to central and peripheral nervous system injuries, burning effects, musculoskeletal effects and ophthalmic effects.

This natural electric discharge has also been responsible for damaging the physical infrastructures of world. Transmission and communication towers, transmission lines and tall physical structures including residential houses and monuments are more vulnerable to lightning activities [12]. Disturbance in the human activities like aviation, outdoor sports and repairable as well as irreparable damages are other most common effects that lightning can pose.

## 1.5 Brief history of scientific study of lightning

Although, lightning was believed to have existed even before the evolution of life on the earth (around 3 billion years ago) and the human being was curious about this phenomenon, scientific experiment was started only in mid 18$^{th}$ century. Experiments proposed and performed by Benjamin Franklin gave birth to lightning as a topic of a research in the scientific world. 'The Sentry Box' experiment proposed by Benjamin Franklin and performed at Marly-la-Ville in 1752 was the formal beginning of scientific research of lightning [2]. After a month of 'The Sentry Box' experiment, Benjamin himself performed the famous 'Kite' experiment in Philadelphia. In "Kite" experiment he observed sparks to jump from a key attached to a kite string to knuckles of his hand. From these experiments Benjamin concluded that "clouds of thunderstorm are most commonly electrically charged"

There is no doubt, the finding of the Benjamin's experiments was important in the field of lightning but the use of lightning photography on a moving film by Hoffert in 1889 was the actual incident that started the scientific progress [2]. Pockels in 1897 measured the lightning current and analyzed the induced magnetic field for the first time [21]. Wilson started to use electric field measurements to evaluate thunderstorm charges involved in lightning discharges. A strong impetus was given to lightning research in the second decade of last century by the needs of electricity supply during thunderstorm periods. Technical necessity thus led to scientific research [13].

At present the experimental methods have become more advanced. Investigations are being carried out with rockets, high-altitude airplanes and spacecrafts. Rocket-triggered lightning research has been an important tool for close-up investigation. Nowadays, the research works on lightning and thunderstorm as a part of atmospheric physics are common. The research is mainly focused to find the correlation of lightning with the global climate, ecology, temperature, vegetation, precipitation etc. Lightning protection system are also being introduced as a gift from science in order to minimize the damage from the disastrous natural process.

In future the Lightning Mapper program has planned to place a sensor in geostationary orbit. This sensor has capacity to map lightning discharges continuously during both day and night, with a special resolution of 10 km.

## 1.6 Classification of ecological zones in Nepal

In Nepal, if ecological maps attempt to portray all recognized differences in landscape characteristics including slope, elevation, total rainfall and its distribution, soil type, micro climate, associated with mature phase and early phase vegetation no planner will be able to grasp how they can be used and certainly no layman can understand them. The ecological classification therefore has to differentiate the major differences only, while allowing for variation within each class [14].

Nepal lies just outside of the tropics in the global climatic zonation. However, bioclimatic tropicality extents into it up to an elevation of 1,000 m altitude. For a mountain country like Nepal altitudinal limits are most convenient to define ecological zones or life zones [15]. On the basis of altitudes from sea level, Nepal has been divided into seven ecological zones. Temperature of these ecological zones goes on decreasing with the increase of altitude. Seven ecological zones of Nepal are as follows:

- Lower Tropical zone: It extends in between (70-300) m altitude from sea level and is the hottest ecological zone of Nepal.

- Upper Tropical zone: This zone lies in the altitudinal limit of (300-1000) m which is below Sub-tropical zone and above Lower Tropical zone.
- Sub-tropical zone: It lies above Upper Tropical Zone up to 2000 m.
- Temperate zone: It is in between Sub-tropical and Sub-alpine zones and in the altitudinal range of (2000-3000) m.
- Sub-alpine zone: This zone extends in between (3000-4000) m altitude from sea level.
- Alpine zone: Alpine zone is the second highest zone in terms of altitude from sea level among ecological zones of Nepal and lies in the altitudinal limit of (4000-5000) m.
- Trans-Himalayan zone: This zone lies above the altitude of 5000 m from sea level and is the one with lowest temperature among seven ecological zones.

Other two ecological zones that are used for the precise study of lightning activities are Water Body and Nival zones. Nival zones do not have the potential of vegetation.

J. F. Dobremez [14], a French researcher, who was the main author of the hard copy maps made in the 1970's and 1980's, sketched the ecological hard copy map of Nepal too[14]. Later on, this hard copy map was used to create digital map which is now used in Geographic Information System (GIS).

## 1.7 Ecological zones and lightning

The thundercloud lightning shows a variety of different characteristics depending on the variability of the size of thundercloud, which in turn depends on the latitude, topography, season and type of storm [5]. Global lightning activity varies from one region to another with the variation in the Earth's climate. The climate of any ecological zone depends upon the amount of radiation obtained from the sun. And the radiations received from sun vary according to the latitudes. As basically different ecological zones of same country are in different latitudes, they have dissimilar climate among them. Thus, it is predictable that the number of lightning strikes varies from one ecological zone to another.

Furthermore, the statistics of the lightning distribution around the world reveals that the intensity and polarity of lightning in thunderstorms are affected by the parameters like surface temperature, water vapor, the troposphere lapse rate and the aerosol loading [3]. These parameters differ in between the ecological zones which supports the evidences of unequal lightning pattern among these zones.

Recent studies continue to show the high positive correlation between surface temperatures and lightning activity [16]. Among the tropical land masses Africa, South America and Southeast Asia, they have the rank of first, second and third consecutively for having greater lightning strikes. They are also called lightning chimneys.The reasons for Africa's strongly continental character and lightning dominance have been attributed to surface characteristics and to the effects of aerosol [17]. Hence warmer ecological zones invite more lightning.

In context of lightning in Nepal, the investigation by Baral and Mackerras in 1992 [18] is worthy to mention. They studied the lightning occurrence characteristics with a flash counter network for a total of 21 months (March 1987–November 1988) in the Kathmandu Valley. Their results indicate that when the lightning activity starts in March, it intensifies quickly reaching its peak in May, while in June the activity decreases rapidly as the monsoon season starts [18]. In this study, Nepal has been divided into nine ecological zones in order to observe the lightning pattern.

## 1.8 Lightning and climate change

Global warming has direct relationship with lightning. From the observations performed on different time scales, we observe a positive relationships between temperature and lightning with lightning increasing anywhere from 10-100 % for every one degree surface warming [16]. Future climate change could have significant repercussions on two related natural hazards: lightning and forest fires [19]. Lightning is predicted to increase by 50% by 2050.

In return, lightning can also make impact on earth's climate. The nitrogen oxides produced during the lightning will assist in minimizing the amount of ultraviolet radiations and other harmful rays in the troposphere by forming ozone, a strong green house gas. But, it also enhances the global warming with its green house effect. As heat, moisture and the wind current are redistributed during thunderstorms and lightning there is change in the climate.

## 1.9 Aim of study

Although having a lot of possibilities and areas for study, research on topics of lightning is in the initial phase in Nepal. Continuing research in this field can be a motivational point to those who have interest in same topic. Available data have shown that Nepal bears varieties in the numbers of lightning strikes in its different ecological zones. Why does a country have different lightning densities among its regions within its territory and what are the factors that could be contributing to give such statistics? The main aim of this study is to search for the answers of these questions by going through the research works that have been done in the related field and by analyzing the data available.

"The process of charge separation in clouds for lightning to origin" is still debatable but no one can question the fact that lightning has been responsible for loss of lives and physical destructions. So, a protective system from lightning is a necessity. The observations and results obtained from this study can emphasize to take steps toward establishing protective systems from lightning. The results of this study can be more than useful to the authorities to differentiate between the areas which are more prone to lightning and those areas which bear less number of strikes. And, hence the knowledge can be availed for the necessary personal safety and structural protection measures.

Furthermore, the distribution of lightning activity over different ecological zones is of much importance to the scientists trying to understand the influence of ground on the lightning activity.

# Chapter 2

## Instrumentation and methodology

The lightning strikes over earth have continuously been monitored by using different sensors. In the present study, we have used the strike data obtained from Global Lightning Network (GLN). One of the sensors of the network been installed on the roof top of Nepal Academy of Science and Technology (NAST) and is the only sensor in Nepal. The strike data obtained from the GLN were analysed with the help of ArcGIS program developed by Environmental Systems Research Institute (ESRI). The ecological map is obtained from International Centre for Integrated Mountain Development (ICIMOD).

GLN is an advanced lightning detection network providing high quality real-time and archive lightning stroke data to partners throughout the world. GLN which can be considered as boon of technology given to humans provided the required data of lightning stroke for this study. This network has become more than efficient in precise detection and in supplying information about lightning activities all over the world so that necessary steps can be taken to minimize the loss of property and save lives from this natural disaster in vulnerable areas.

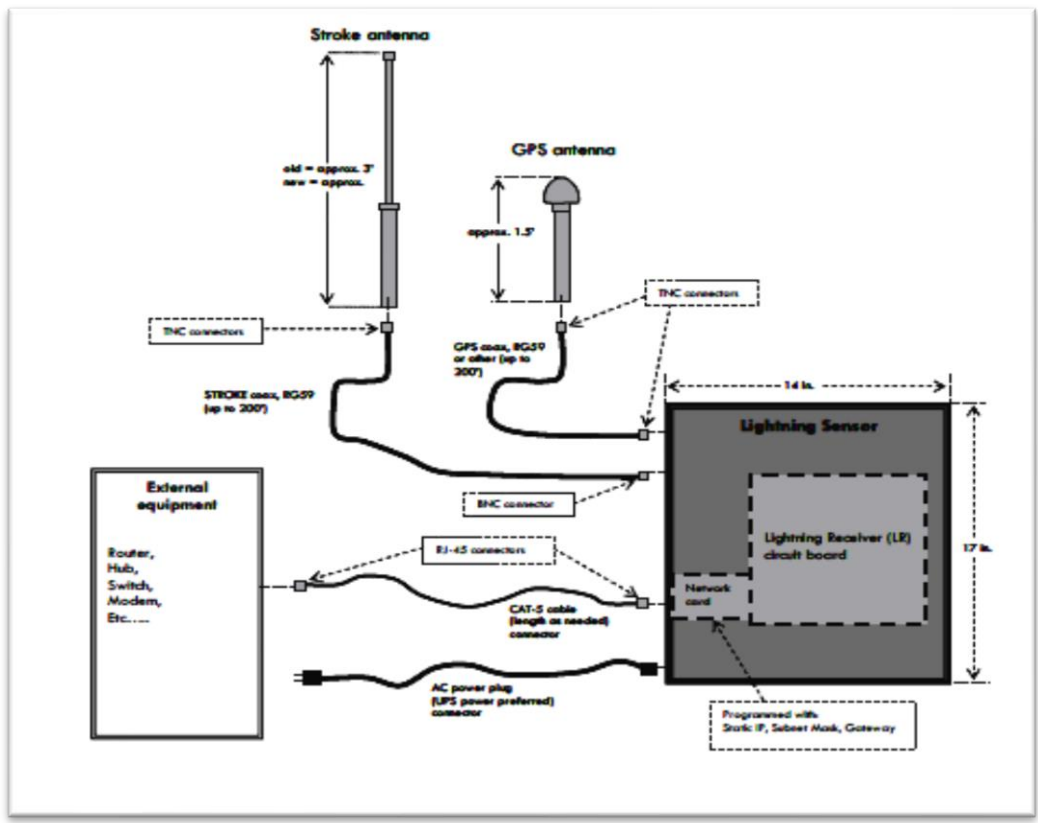

Figure 2.1: Schematic diagram of sensor with stroke and GPS antenna. Adapted from "Study of effect of climate change on lightning activity in Nepal" [21].

GLN uses Time Difference of Arrival (TOA) Systems Precision Lightning Sensors in order to detect and analyze the lightning activities precisely. Each sensor system has a sensor receiver chassis accompanied with Global Positioning System (GPS) antenna and stroke antenna. In a time-of-arrival based system, timing is an important part of the receiver and the receiver uses GPS timing as a reference. When the raw data are forwarded to TOA Systems Central Analysis Processor (CAP), they are analyzed over there to produce transformable solutions within 10 seconds or less. CAP computes and displays real time lightning location information. All the lightning activities are archived by (Weather Services International) WSI Corporation and made them available to the host partners.

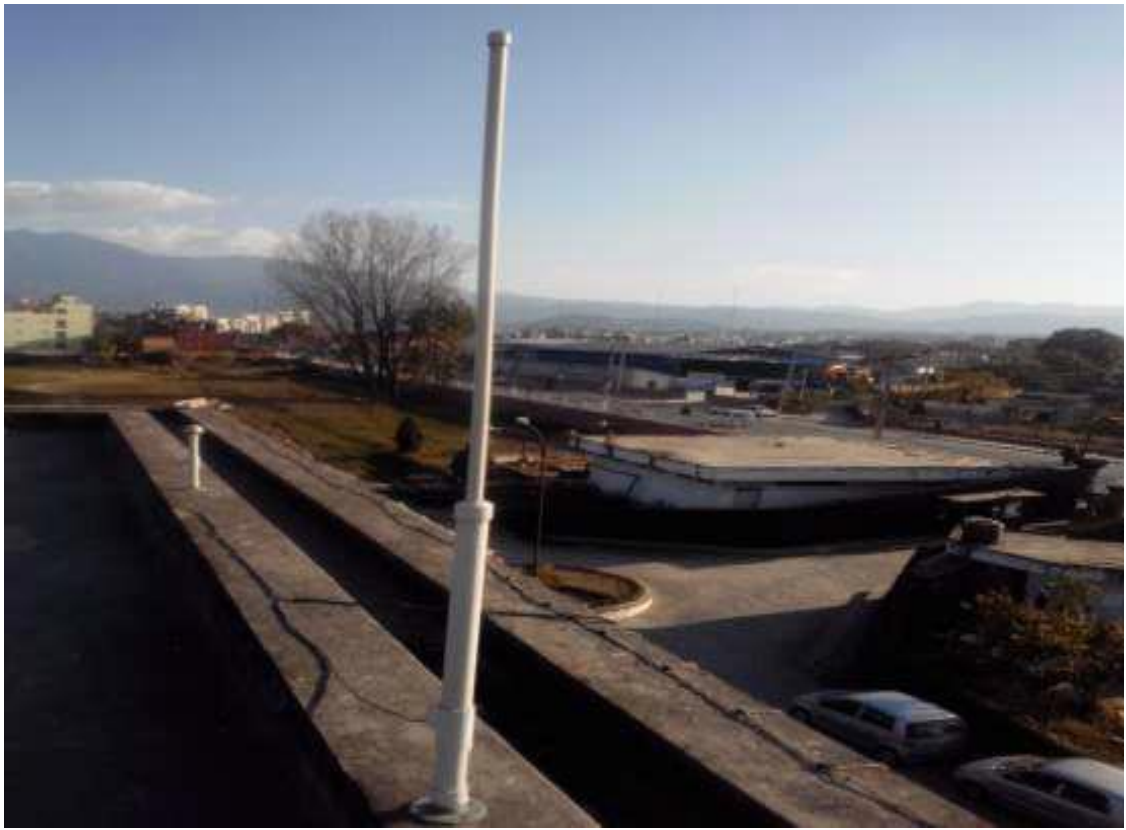

Figure 2.2: Photograph of stroke antenna and GPS antenna on rooftop of NAST.

GPS is an operational system, providing users worldwide with twenty-four hour a day precise position in three dimensions and precise time traceable to global time standards [20]. It consists of a constellation of 24 satellites that continuously orbit the Earth. Each GPS satellite has on board several atomic clocks that are precisely synchronized to Coordinated Universal Time provided by the U.S. Naval Observatory (USNO). In order to acquire the signals, GPS antenna should be mounted on the roof or window for clear view of the sky.

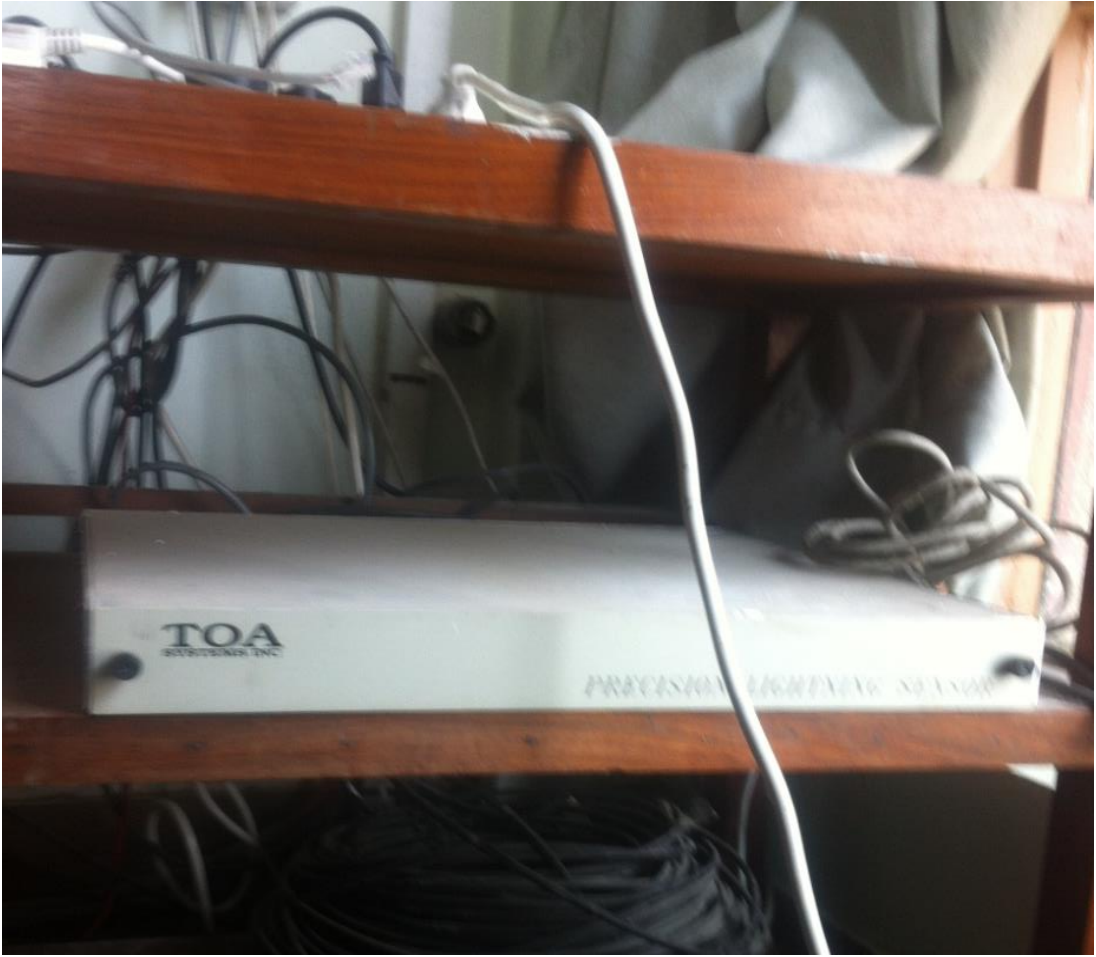

Figure 2.3: Photograph of TOA sensor at NAST.

After having the data of lightning activities of all over Nepal and nearby locations, they were projected, clipped and finalised to collect necessary information with the help of software ArcMap which is one of the part of software ArcGIS. Data from December 2011 to February 2015 excluding time period of September 2013 to September 2014 were analyzed. We had no other choice than to exclude data for the time period mentioned because the system in Nepal to collect raw data of lightning was in the condition to be repaired. The available data were plotted by separating them into four parts as per the four seasons of the year. Four seasons used are listed as follows:

(1) Winter season (December, January and February)
(2) Pre-monsoon season (March, April and May)
(3) Monsoon season (June, July and August)
(4) Post-monsoon season (September, October and November)

Ecological map of Nepal with nine ecological zones has been incorporated in ArcGIS and the number of lightning strikes assigned to each ecological zone along with the area of ecological zones was obtained. Thereafter, density of lightning in the required zone was calculated as:

$$\text{Lightning Density} = \frac{\text{Number of lightning strikes}}{\text{Area in square kilometer}} \quad \ldots\ldots\ldots\ldots\ldots (2.1)$$

Thus, obtained different values of number of lightning strikes and densities of lightning for different ecological zones were presented with line graph for each season. Also, each ecological zone was indexed with different colour so that the dominancy of lightning activities in those regions can be analysed properly with the help of ArcGIS.

Although, Nepal has been divided into seven ecological zones on the basis of altitude from sea level, map of Nepal separated into nine ecological divisions was used for detail and more precise study as Nival Zones and Water Body may play significant role on lightning activities.

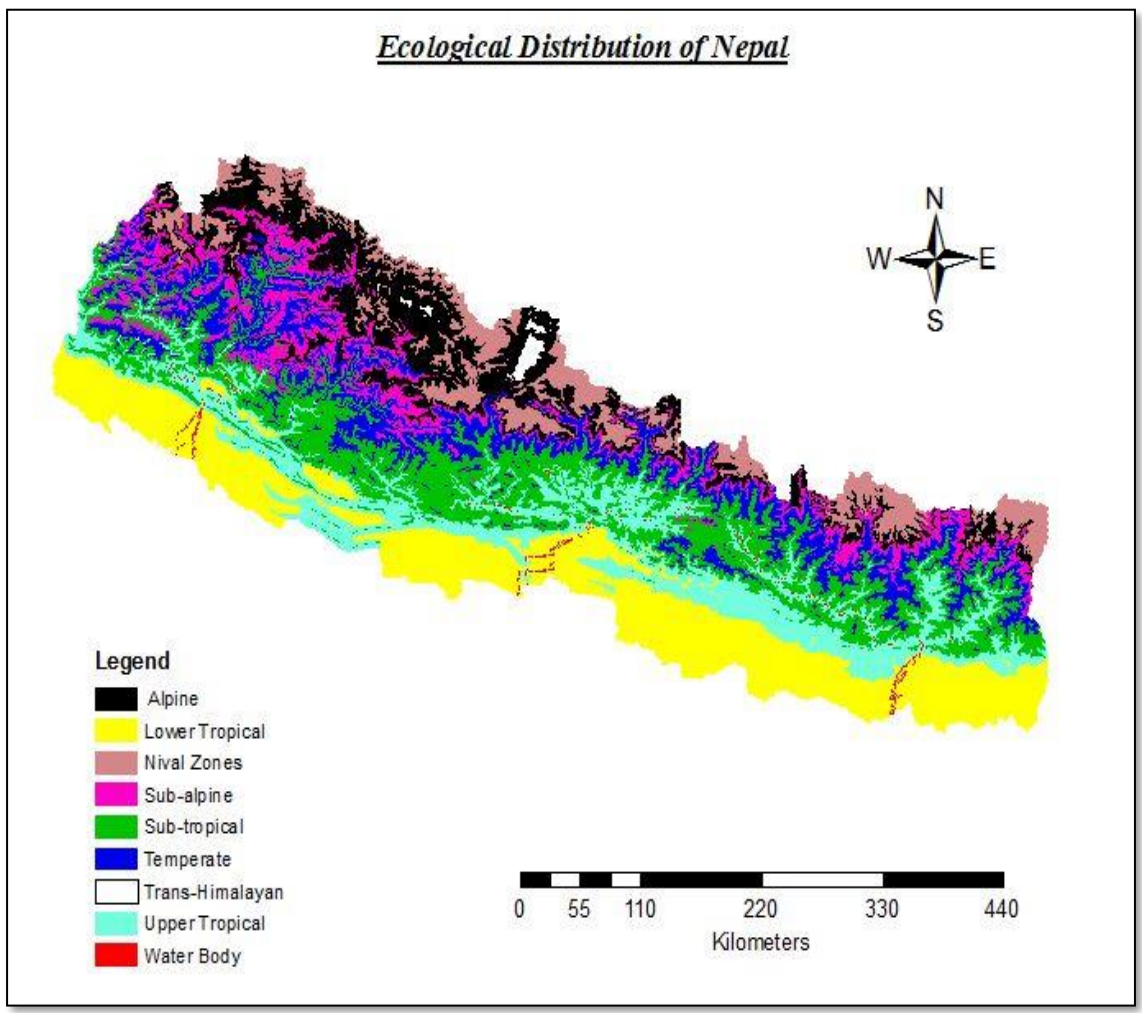

Figure 2.4: Ecological division of Nepal.

Seven ecological zones of Nepal on the basis of altitude are tabulated as:

Table 2.1: Ecological zones of Nepal with altitude and area covered in percentage.

| S.N | Ecological zone | Altitude (m) | Area Covered (% of Nepal) |
|-----|-----------------|--------------|---------------------------|
| 1 | Lower Tropical | 70-300 | 18 |
| 2 | Upper Tropical | 300-1000 | 18 |
| 3 | Sub-tropical | 1000-2000 | 22 |
| 4 | Temperate | 2000-3000 | 12 |
| 5 | Sub-alpine | 3000-4000 | 9 |
| 6 | Alpine | 4000-5000 | 8 |
| 7 | Trans-Himalayan | Above 5000 | 8 |

Other two ecological zones over which the lightning activities were studied are Nival Zones and Water Body.

**Measurement, parameter and unit**

Position of sensor in Nepal: NAST, Khumaltar, Lalitpur
Latitude: $27.65^0$ N
Longitude: $85.32^0$ E
Altitude: 1.38 km (average sea level)
Lightning Density: Number of lightning strikes per square kilometer
Time: Days, Months, Seasons (Pre-monsoon, monsoon, post-monsoon and winter)

# Chapter 3

# Observation

**Distribution of lightning activity over different ecological zones**

Lightning strikes occurring over the globe have continuously been monitored by the WSI's Global Lightning Network (GLN) system. The electric field from lightning being sensed by the stroke antenna located by the coordinated GPS system are being processed and archived by the WSI's CAP. The data so archived have further been processed and analyzed in this study with the help of ArcGIS. We have used ecological map of Nepal (shown in figure 2.4) in order to clip and make spatial analysis of the available data of lightning using ArcGIS. After finding number of lightning strikes for different zones, the densities of lightning strikes in units of per square kilometer are calculated for each regions using equation (2.1). The density so obtained were tabulated and represented through line graph in this chapter. The area of each ecological zone used for calculation is given in the table 3.1.

Table 3.1: Ecological zones of Nepal with area covered in square kilometer.

| S.N | Ecological zones | Area ($km^2$) |
|---|---|---|
| 1 | Alpine | 17088 |
| 2 | Lower Tropical | 25153 |
| 3 | Nival Zones | 11061 |
| 4 | Sub-alpine | 14326 |
| 5 | Sub-tropical | 31707 |
| 6 | Temperate | 20915 |
| 7 | Trans-Himalayan | 3224 |
| 8 | Upper Tropical | 23911 |
| 9 | Water Body | 638 |

## 3.1 Distribution of lightning activity over different ecological zones of Nepal for different seasons

Lightning activities occurring all over Nepal were observed by separating the available data into four different seasons namely winter, pre-monsoon, monsoon and post-monsoon respectively. Observed values and calculations under the heading of four seasons are as follows.

**(A) Winter**

**(a) Winter 2012**

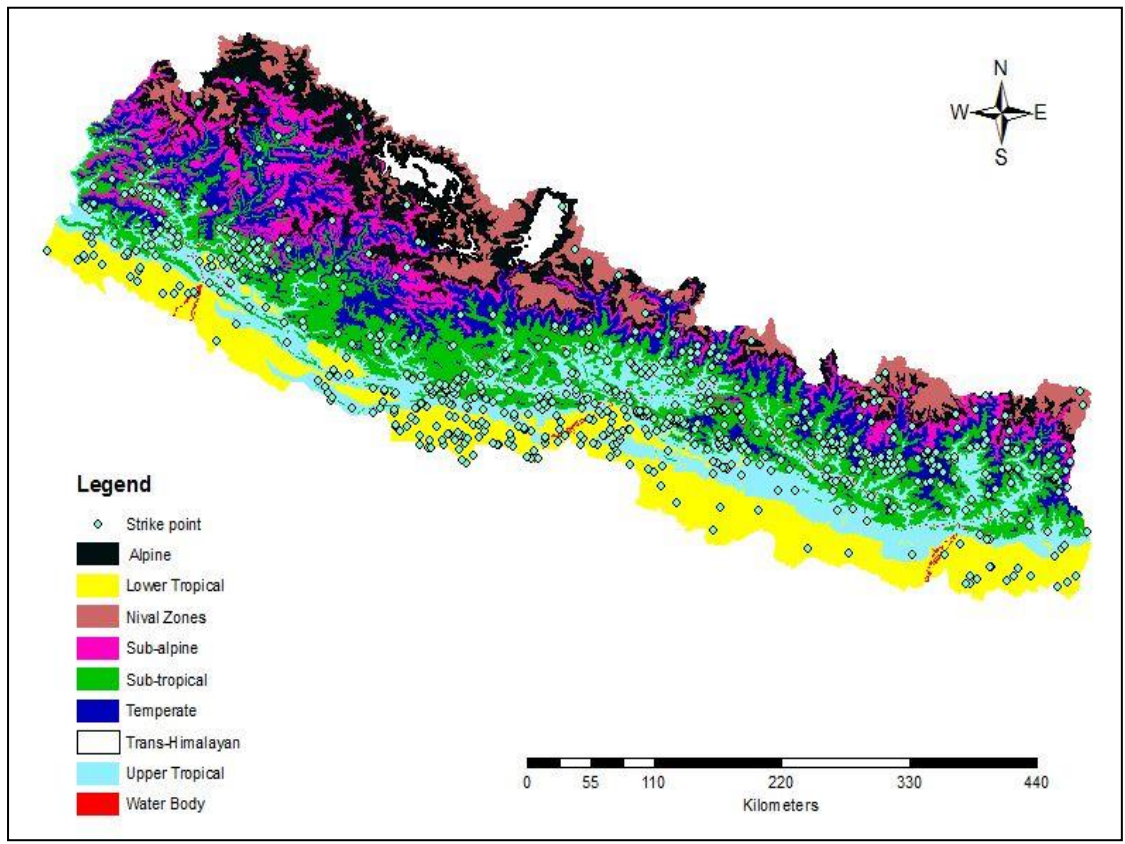

Figure 3.1: Ecological map of Nepal with lightning strikes during winter 2012.

Table 3.2: Lightning activity during winter 2012 over different ecological zones of Nepal.

| S.N | Ecological zones | Number of lightning strikes | Density of lightning strikes (per square kilometer) |
|---|---|---|---|
| 1 | Alpine | 12 | $0.07 \times 10^{-2}$ |
| 2 | Lower Tropical | 144 | $0.57 \times 10^{-2}$ |
| 3 | Nival Zones | 12 | $0.10 \times 10^{-2}$ |
| 4 | Sub-alpine | 34 | $0.23 \times 10^{-2}$ |
| 5 | Sub-tropical | 216 | $0.68 \times 10^{-2}$ |
| 6 | Temperate | 106 | $0.50 \times 10^{-2}$ |
| 7 | Trans-Himalayan | 1 | $0.03 \times 10^{-2}$ |
| 8 | Upper Tropical | 158 | $0.66 \times 10^{-2}$ |
| 9 | Water Body | 5 | $0.78 \times 10^{-2}$ |

From the map obtained via ArcGIS, by incorporating the strike data, we were able to view the facts related to lightning activity for different zones during the time period concerned. During winter 2012 i.e. December 2011 to February 2012, Nepal experienced total of 688 lightning strikes. Out of which Sub-tropical zone experienced the most with around 31 percent of strikes whereas Trans-Himalayan was the one to have least number of strikes with only one strike for the time period under consideration. Analyzing on the basis of altitude, (table 2.1 for reference), we can see that those zones which are in high altitude from sea level have less number of strikes and zones which are in less altitude from sea level have high number of strikes.

Dealing with the densities of lightning strikes, it was found that Water Body and Sub-tropical zones are in the first and second place respectively to have higher densities of strikes and Alpine and Trans-Himalayan zones were in second last and last position respectively. Furthermore, details of lightning activity during winter 2012 for other ecological zones are listed in table 3.2 above.

In figure 3.2 values of number of lightning strikes and density of lightning strikes $\times$ 10000 over ecological zones were plotted in the line graph for quick interpretation. Value of density of strikes was multiplied by 10000 in order to bring the value in the range of number of lightning strikes so that they can be plotted and analyzed at once in the same graph.

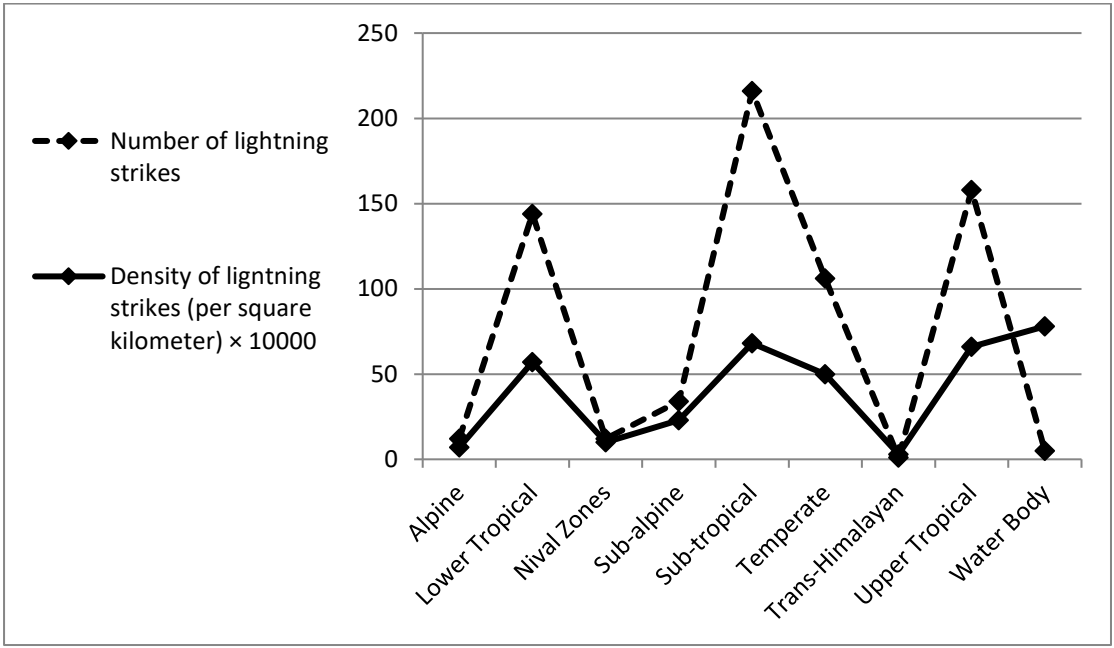

Figure 3.2: Lightning activity during winter 2012 over different ecological zones of Nepal.

**(b) Winter 2013**

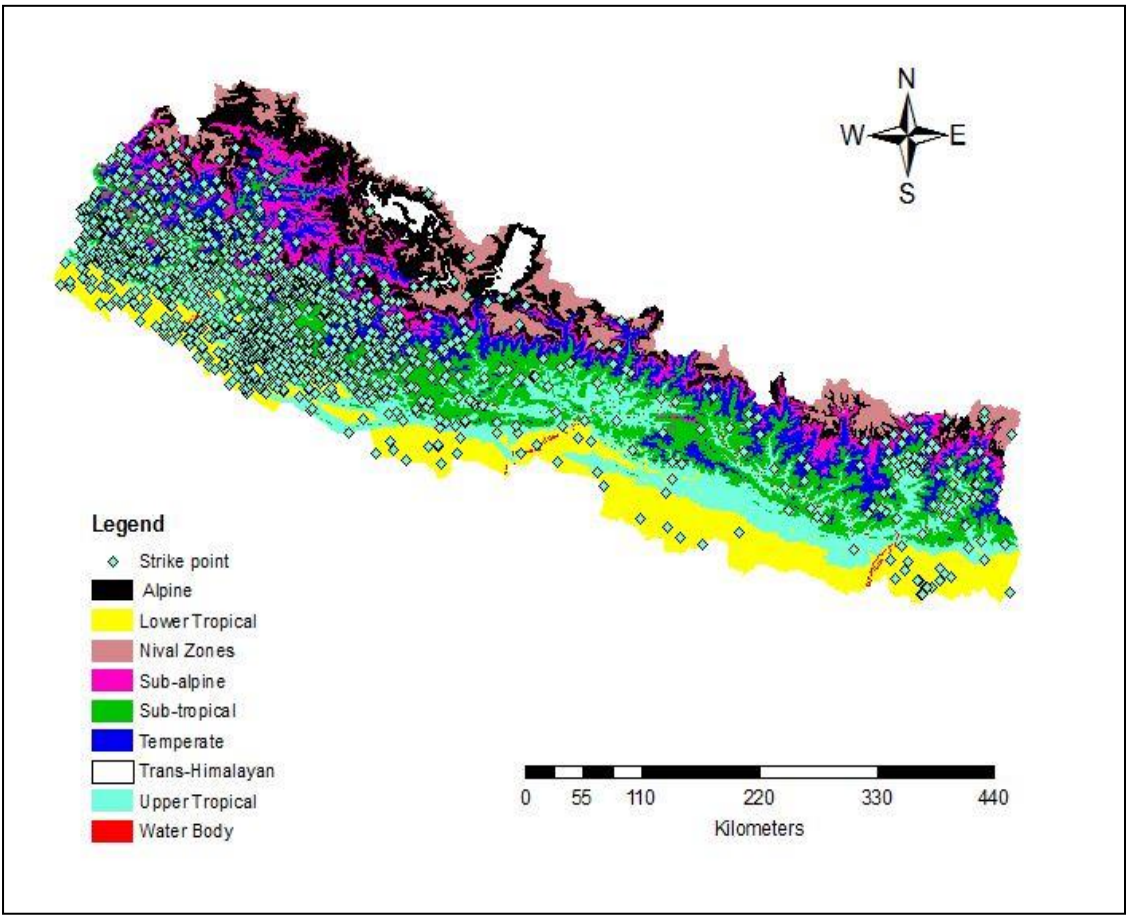

Figure 3.3: Ecological map of Nepal with lightning strikes during winter 2013.

Table 3.3: Lightning activity during winter 2013 over different ecological zones of Nepal.

| S.N | Ecological zones | Number of lightning strikes | Density of lightning strikes (per square kilometer) |
|---|---|---|---|
| 1 | Alpine | 39 | $0.23 \times 10^{-2}$ |
| 2 | Lower Tropical | 192 | $0.76 \times 10^{-2}$ |
| 3 | Nival Zones | 8 | $0.07 \times 10^{-2}$ |
| 4 | Sub-alpine | 121 | $0.84 \times 10^{-2}$ |
| 5 | Sub-tropical | 425 | $1.34 \times 10^{-2}$ |
| 6 | Temperate | 232 | $1.11 \times 10^{-2}$ |
| 7 | Trans-Himalayan | 0 | 0 |
| 8 | Upper Tropical | 214 | $0.89 \times 10^{-2}$ |
| 9 | Water Body | 4 | $0.63 \times 10^{-2}$ |

We were able to draw table 3.3 from the map obtained from ArcGIS for winter 2013. It is clearly seen from table 3.3 that Sub-tropical zone exceeded all other zones with 425 out of 1235 strikes and Trans-Himalayan zone was the one to have least value as it did not have any strike during winter 2013 (December 2012 to February 2013). Positions of ecological zones with highest and least number of strikes were found to be same for both 2012 and 2013 winters.

Moving on to the density of lightning strikes, Sub-tropical zone is found to have maximum value of $1.34 \times 10^{-2}$ / km$^2$ while Trans-Himalayan zone is found to have minimum value of 0. For winter 2013, Sub-tropical, Temperate, Upper Tropical zones were in the first, second and third place respectively for densities of lightning while Alpine, Nival and Trans-Himalayan zones were in the seventh, eighth and ninth place respectively. Sub-alpine, Lower Tropical and Water Body zones fell in the fourth, fifth and sixth place respectively.

Table 3.3 is represented with the help of graph below in figure 3.4 after multiplying the column of density of lightning strikes by 10000. We can see from graph that although Water Body zone has the value of number of lightning strikes in the lower side among the ecological zones, it has got its value in higher side for density of lightning strikes.

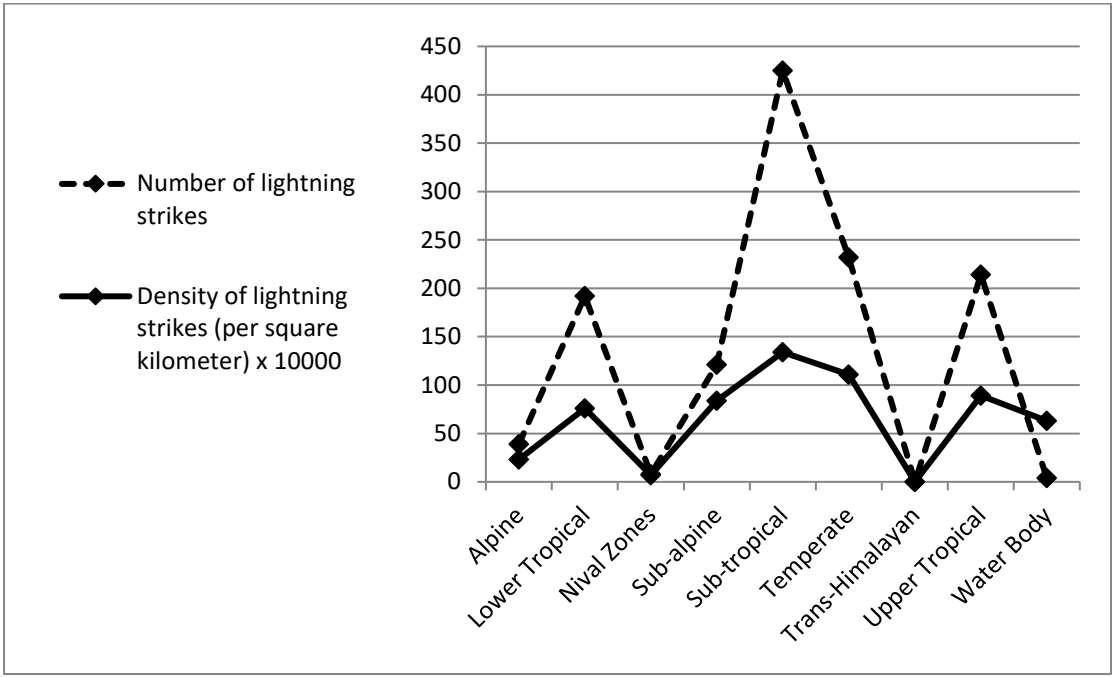

Figure 3.4: Lightning activity during winter 2013 over different ecological zones of Nepal.

**(c) Winter 2015**

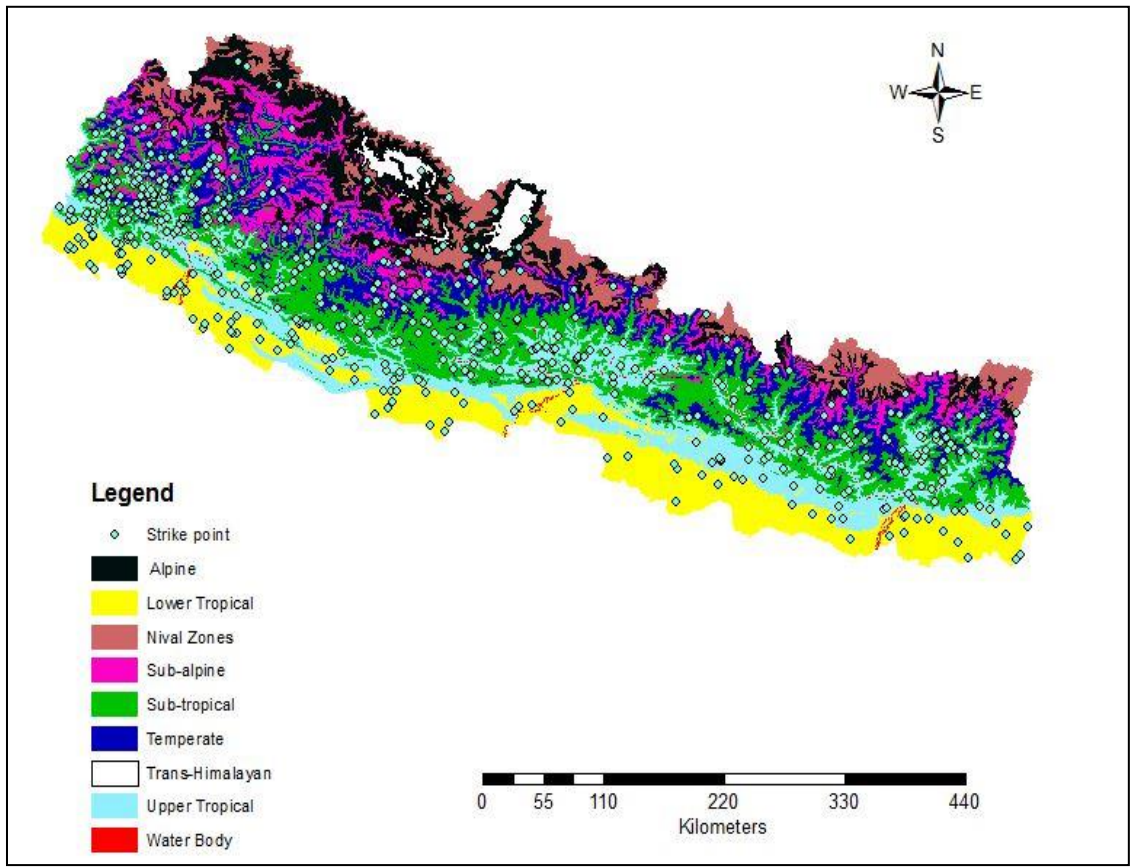

Figure 3.5: Ecological map of Nepal with lightning strikes during winter 2015.

Table 3.4: Lightning activity during winter 2015 over different ecological zones of Nepal.

| S.N | Ecological zones | Number of lightning strikes | Density of lightning strikes (per square kilometer) |
|---|---|---|---|
| 1 | Alpine | 17 | $0.10 \times 10^{-2}$ |
| 2 | Lower Tropical | 87 | $0.35 \times 10^{-2}$ |
| 3 | Nival Zones | 13 | $0.12 \times 10^{-2}$ |
| 4 | Sub-alpine | 51 | $0.36 \times 10^{-2}$ |
| 5 | Sub-tropical | 163 | $0.51 \times 10^{-2}$ |
| 6 | Temperate | 80 | $0.38 \times 10^{-2}$ |
| 7 | Trans-Himalayan | 2 | $0.06 \times 10^{-2}$ |
| 8 | Upper Tropical | 107 | $0.45 \times 10^{-2}$ |
| 9 | Water Body | 1 | $0.16 \times 10^{-2}$ |

Winter season of 2015 (December 2014 to February 2015) is found to have the minimum number of strikes among the winters under study. Total number of lightning strikes for this season was found to be 521. As during winter seasons of 2012 and 2013, Sub-tropical zone left behind all other ecological zones for having maximum number of strikes with 31 percent of total lightning strikes. Water Body zone experienced only one strike and Trans-Himalayan zone is found to have two strikes which are the minimum values for this time period. From table 3.4 it is clear that, the highest value for density of lightning strikes is $0.51 \times 10^{-2}$ / km$^2$ which is of Sub-tropical zone and the lowest value is $0.06 \times 10^{-2}$ / km$^2$ assigned to Temperate zone.

Figure 3.6 depicts lightning activity over the nine different ecological zones of Nepal during winter 2015. This figure clearly shows peaks for three tropical zones which indicate the dominancy in lightning activity by Lower Tropical, Upper Tropical and Sub-tropical zones.

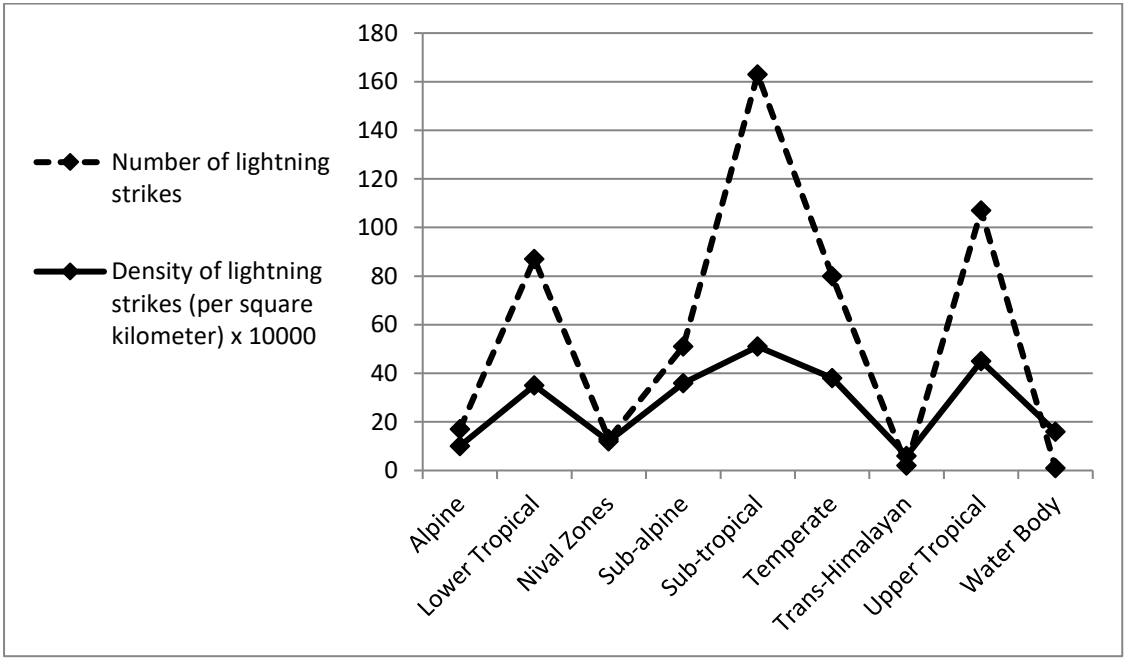

Figure 3.6: Lightning activity during winter 2015 over different ecological zones of Nepal.

**(B) Pre-monsoon**

**(a) Pre-monsoon 2012**

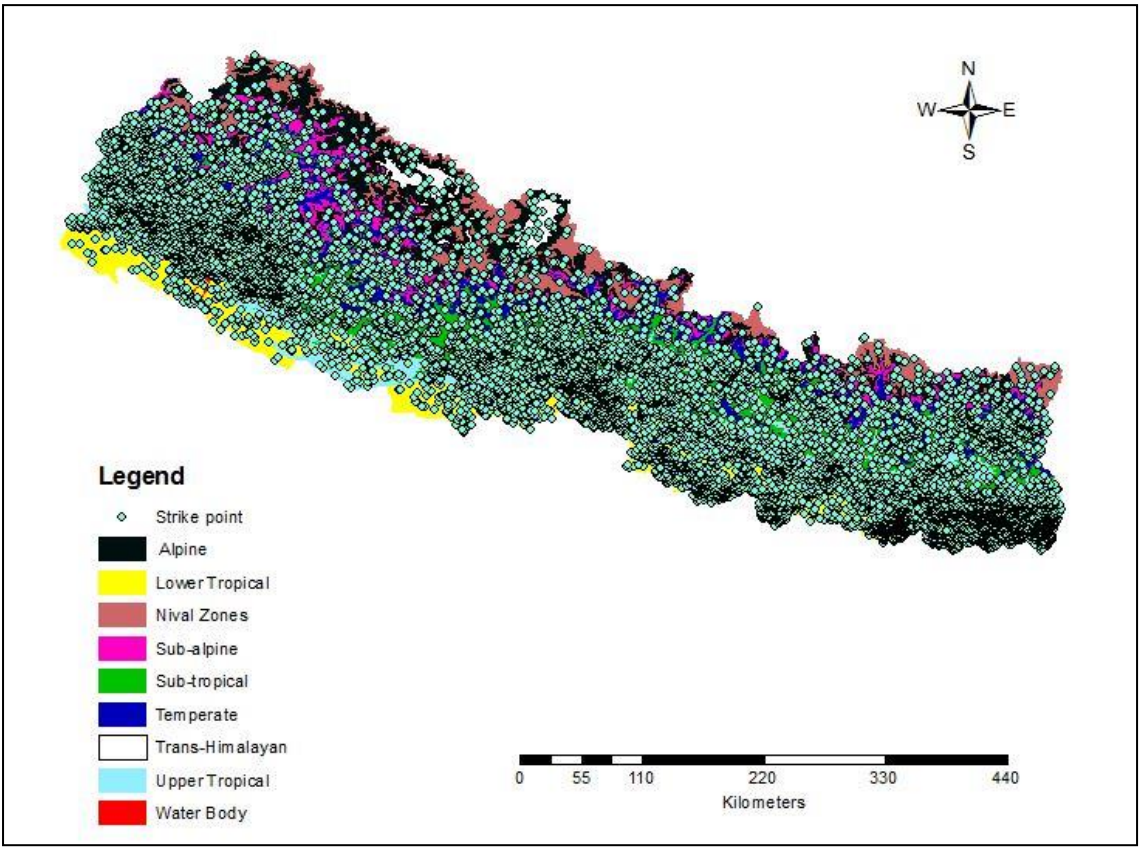

Figure 3.7: Ecological map of Nepal with lightning strikes during pre-monsoon 2012.

Table 3.5: Lightning activity during pre-monsoon 2012 over different ecological zones of Nepal.

| S.N | Ecological zones | Number of lightning strikes | Density of lightning strikes (per square kilometer) |
|---|---|---|---|
| 1 | Alpine | 329 | $1.92 \times 10^{-2}$ |
| 2 | Lower Tropical | 3774 | $15.00 \times 10^{-2}$ |
| 3 | Nival Zones | 142 | $1.28 \times 10^{-2}$ |
| 4 | Sub-alpine | 729 | $5.08 \times 10^{-2}$ |
| 5 | Sub-tropical | 3075 | $9.70 \times 10^{-2}$ |
| 6 | Temperate | 1489 | $7.11 \times 10^{-2}$ |
| 7 | Trans-Himalayan | 50 | $1.55 \times 10^{-2}$ |
| 8 | Upper Tropical | 2947 | $12.32 \times 10^{-2}$ |
| 9 | Water Body | 94 | $14.73 \times 10^{-2}$ |

During pre-monsoon (March, April and May) 2012, 12629 lightning strikes were recorded. Lower Tropical zone and Sub-tropical zone respectively faced 3774 and 3075 strikes which were most for the time period of March to April of 2012. Similarly, the least values were 50 and 94 assigned to Trans-Himalayan zone and Water Body zone. In spite of having the second lowest value in column of number of lightning strikes, Water Body zone has second highest value of $14.73 \times 10^{-2}$/ km² in the column of density of lightning strikes in table 3.5. Lower Tropical zone has the maximum value of density of lightning strikes as $15.00 \times 10^{-2}$/ km² among nine ecological zones while the minimum value of $1.28 \times 10^{-2}$/ km² is of Nival Zones.

Figure 3.8 is a graph plotted on the basis of table 3.5 and can be used to make analysis of lightning activity over ecological zones of Nepal during pre-monsoon 2012. It clearly indicates the dominance of number of lightning strikes by Lower Tropical zone and also represents the dominance of density of lightning strikes by Lower Tropical and Water Body zones.

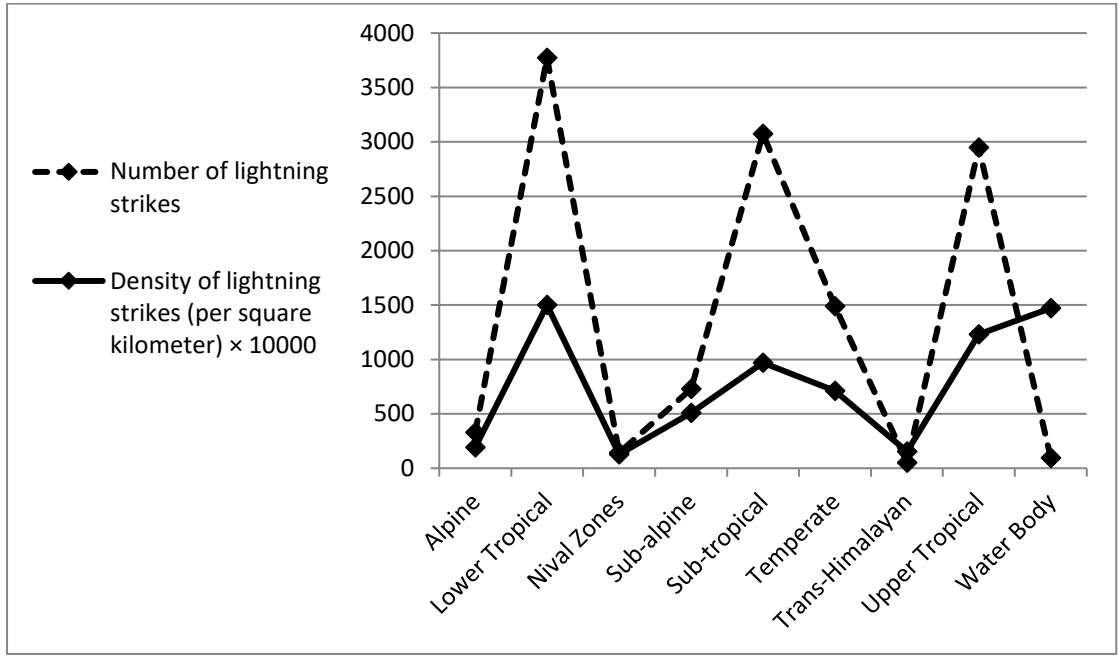

Figure 3.8: Lightning activity during pre-monsoon 2012 over different ecological zones of Nepal.

**(b)  Pre-monsoon 2013**

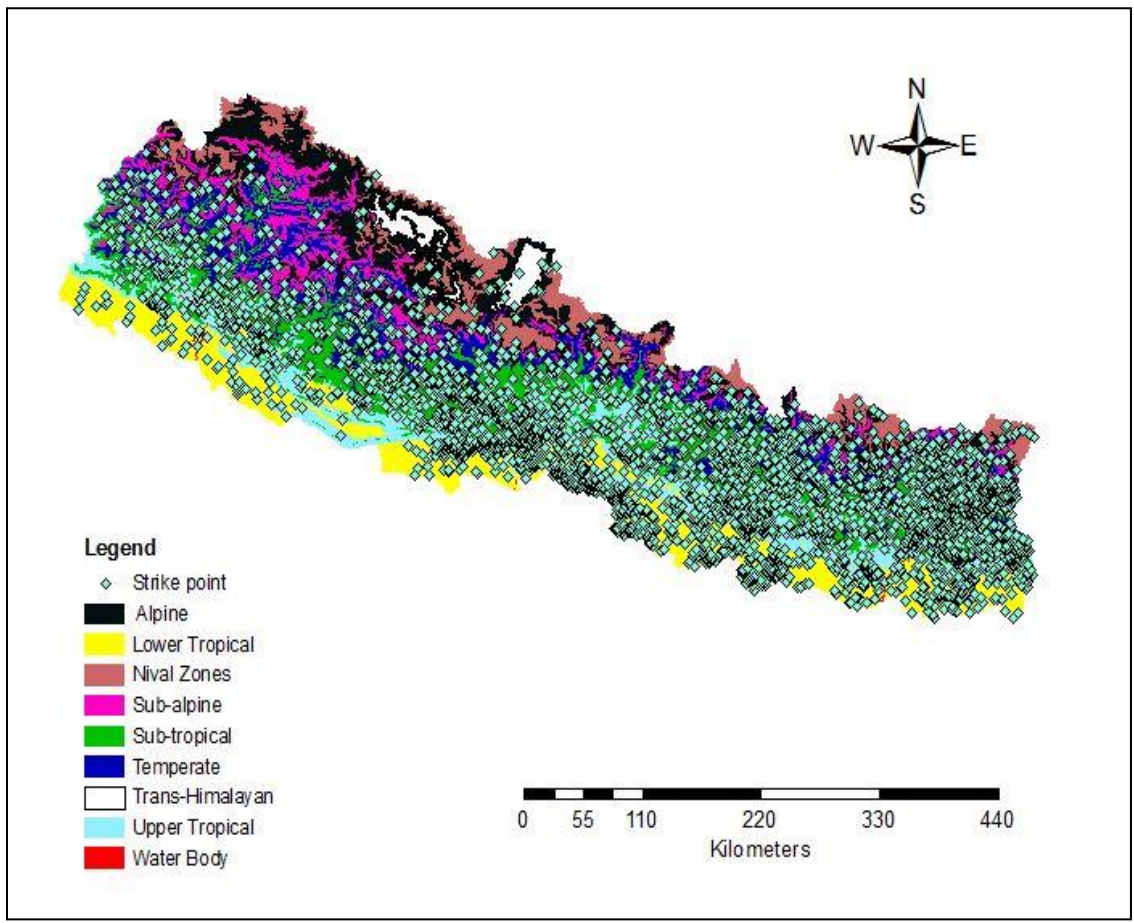

Figure 3.9: Ecological map of Nepal with lightning strikes during pre-monsoon 2013.

**27**

Table 3.6: Lightning activity during pre-monsoon 2013 over different ecological zones of Nepal.

| S.N | Ecological zones | Number of lightning strikes | Density of lightning strikes (per square kilometer) |
|---|---|---|---|
| 1 | Alpine | 111 | $0.65 \times 10^{-2}$ |
| 2 | Lower Tropical | 823 | $3.27 \times 10^{-2}$ |
| 3 | Nival Zones | 56 | $0.50 \times 10^{-2}$ |
| 4 | Sub-alpine | 219 | $1.52 \times 10^{-2}$ |
| 5 | Sub-tropical | 1082 | $3.41 \times 10^{-2}$ |
| 6 | Temperate | 581 | $2.78 \times 10^{-2}$ |
| 7 | Trans-Himalayan | 6 | $0.19 \times 10^{-2}$ |
| 8 | Upper Tropical | 855 | $3.58 \times 10^{-2}$ |
| 9 | Water Body | 20 | $3.13 \times 10^{-2}$ |

After clipping and making spatial analysis of the lightning data with ArcGIS, we had an observation that Nepal experienced a total of 3753 lightning strikes during pre-monsoon (March, April and May) 2013. Most of these strikes struck over Sub-tropical zone with value of 1082 which is about 29 percent of total strikes. The lowest number of strikes recorded is found to be 6 over the Trans-Himalayan zone which is only about 0.16 percent of the total number of lightning strikes. Moving on to the density of lightning strikes, table 3.4 clearly shows that Lower Tropical, Sub-tropical, Upper Tropical and Water Body zones have higher values of $3.27 \times 10^{-2}/\mathrm{km}^2$, $3.41 \times 10^{-2}/\mathrm{km}^2$, $3.58 \times 10^{-2}/\mathrm{km}^2$ and $3.13 \times 10^{-2}/\mathrm{km}^2$ respectively. Densities for Alpine, Nival Zones and Trans-Himalayan have lower values of $0.65 \times 10^{-2}/\mathrm{km}^2$, $0.50 \times 10^{-2}/\mathrm{km}^2$ and $0.19 \times 10^{-2}/\mathrm{km}^2$.

Figure 3.10 is available for quick interpretation of lightning activity over different ecological zones of Nepal during pre-monsoon 2013.

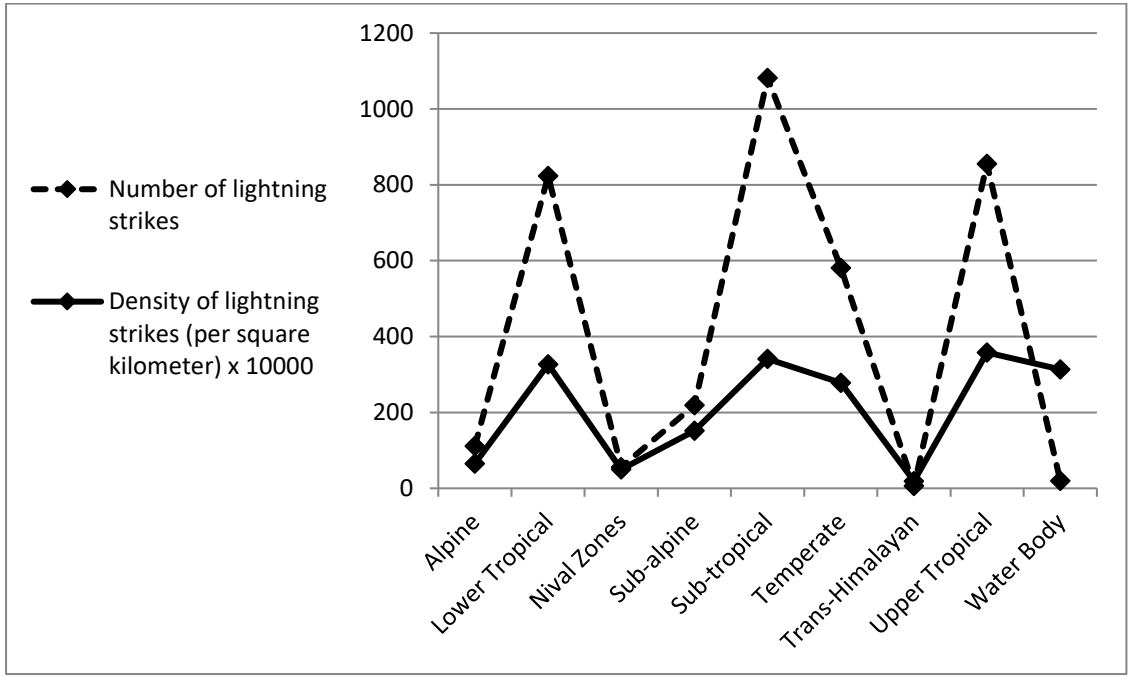

Figure 3.10: Lightning activity during pre-monsoon 2013 over different ecological zones of Nepal.

**(C) Monsoon**

**(a) Monsoon 2012**

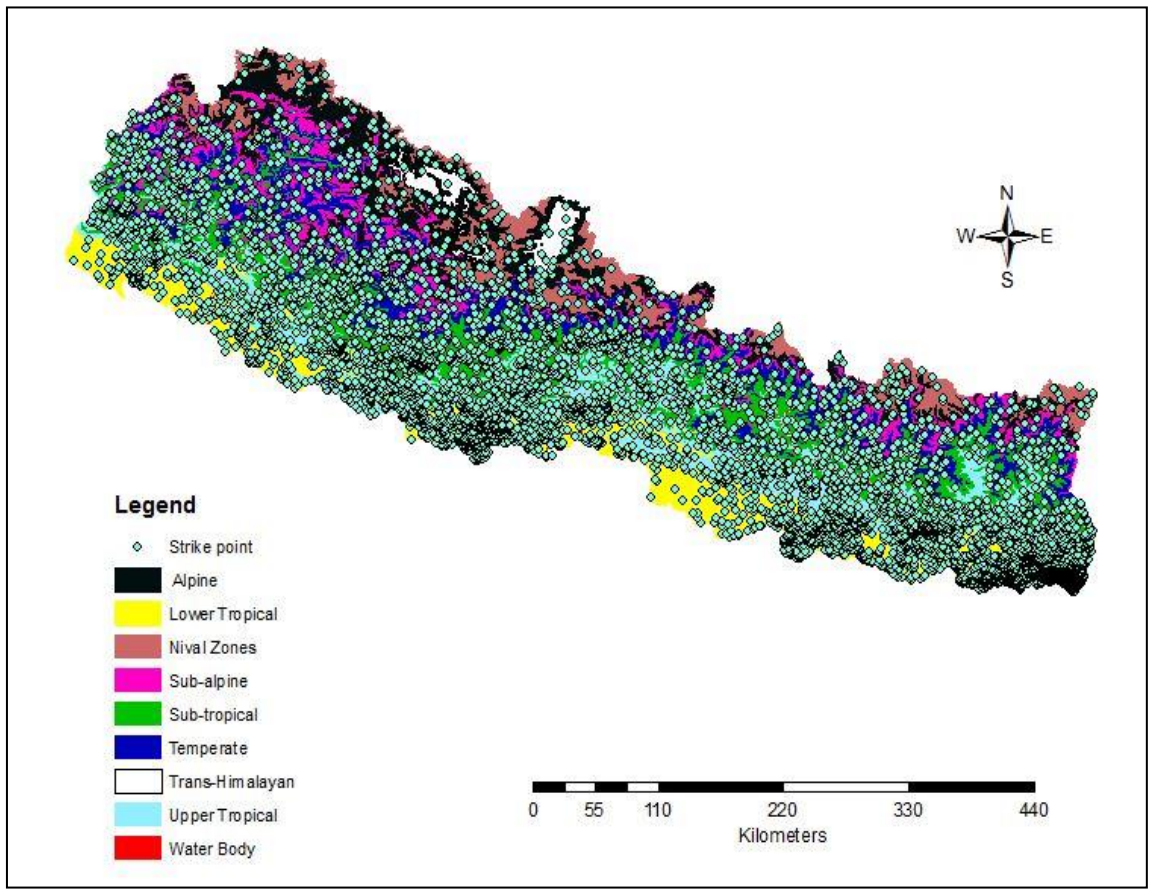

Figure 3.11: Ecological map of Nepal with lightning strikes during monsoon 2012.

Table 3.7: Lightning activity during monsoon 2012 over different ecological zones of Nepal.

| S.N | Ecological zones | Number of lightning strikes | Density of lightning strikes (per square kilometer) |
|---|---|---|---|
| 1 | Alpine | 173 | $1.01 \times 10^{-2}$ |
| 2 | Lower Tropical | 3401 | $13.52 \times 10^{-2}$ |
| 3 | Nival Zones | 118 | $1.06 \times 10^{-2}$ |
| 4 | Sub-alpine | 255 | $1.77 \times 10^{-2}$ |
| 5 | Sub-tropical | 1267 | $3.10 \times 10^{-2}$ |
| 6 | Temperate | 530 | $2.53 \times 10^{-2}$ |
| 7 | Trans-Himalayan | 28 | $0.87 \times 10^{-2}$ |
| 8 | Upper Tropical | 1351 | $5.65 \times 10^{-2}$ |
| 9 | Water Body | 30 | $4.70 \times 10^{-2}$ |

From the map obtained via ArcGIS, incorporating the strike data, we found that monsoon (June, July and August) 2012 received 7153 lightning strikes. About 48 percent of strikes are clustered over Lower Tropical zone as shown in figure 3.11. Observing table 3.7, we can say that the region which ranges from 0 to 300 meter in altitude from sea level i.e. Lower Tropical zone experienced highest number of strikes of 3401 and the region which is farthest from sea level i.e. Trans-Himalayan zone which is above 5000 meter from sea level experienced the least strikes of 28. If we see the densities of lightning strike zones highest and lowest values are same as that for the number of strikes. Lower Tropical zone has the maximum value of $13.52 \times 10^{-2}$ / km$^2$ by far among the ecological zones and the minimum value of $0.87 \times 10^{-2}$ / km$^2$ belongs to Trans-Himalayan zone.

Line graph plotted below in figure 3.12 depicts lightning activity of monsoon 2012 which clearly reports that Lower Tropical has peak values for number of lightning strikes and density of lightning strikes among the ecological zones.

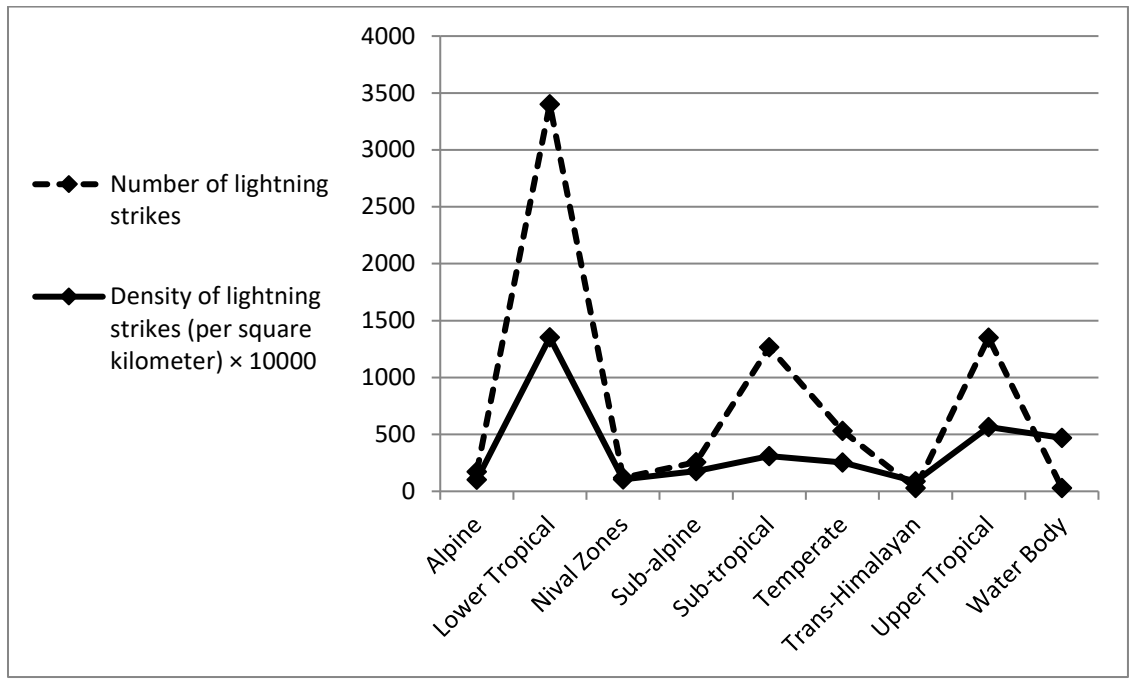

Figure 3.12: Lightning activity during monsoon 2012 over different ecological zones of Nepal.

**(b) Monsoon 2013**

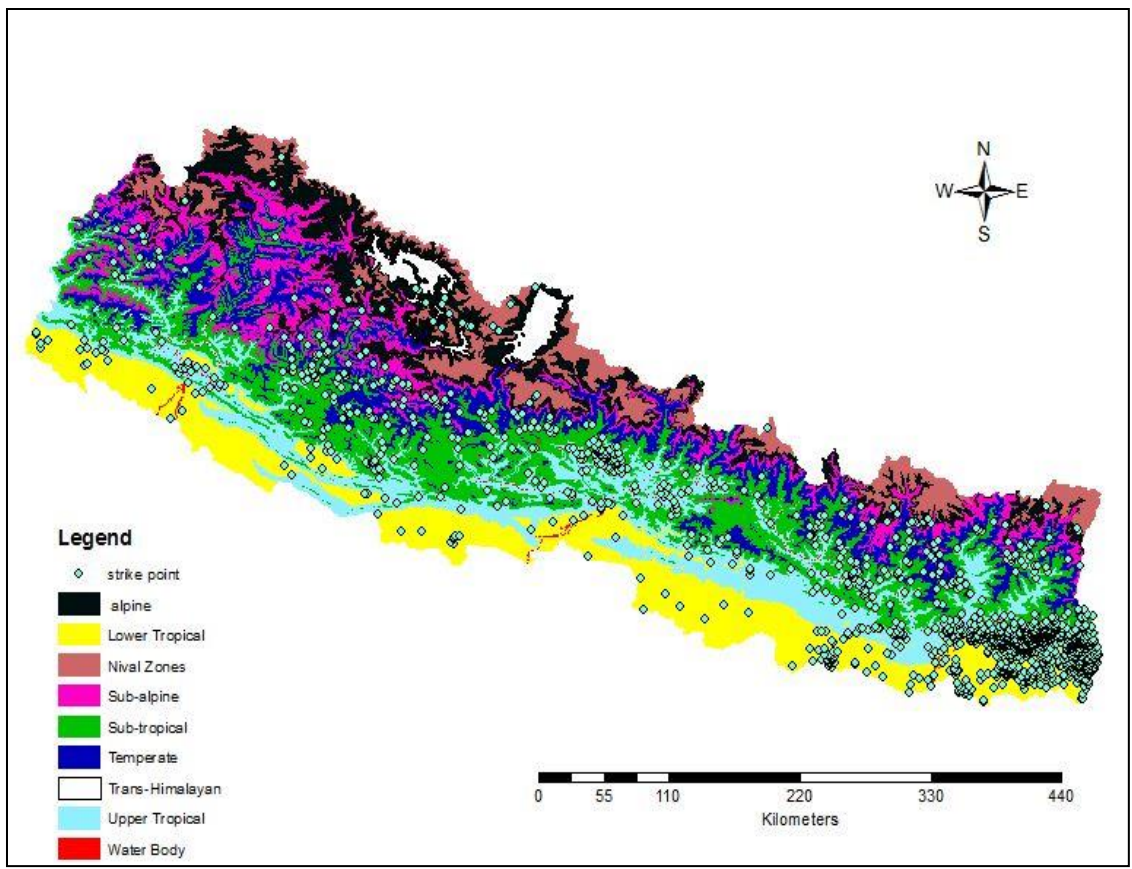

Figure 3.13: Ecological map of Nepal with lightning strikes during monsoon 2013.

Table 3.8: Lightning activity during monsoon 2013 over different ecological zones of Nepal.

| S.N | Ecological zones | Number of lightning strikes | Density of lightning strikes (per square kilometer) |
| --- | --- | --- | --- |
| 1 | Alpine | 21 | $0.12 \times 10^{-2}$ |
| 2 | Lower Tropical | 447 | $1.78 \times 10^{-2}$ |
| 3 | Nival Zones | 16 | $0.14 \times 10^{-2}$ |
| 4 | Sub-alpine | 51 | $0.36 \times 10^{-2}$ |
| 5 | Sub-tropical | 467 | $1.47 \times 10^{-2}$ |
| 6 | Temperate | 127 | $0.60 \times 10^{-2}$ |
| 7 | Trans-Himalayan | 1 | $0.03 \times 10^{-2}$ |
| 8 | Upper Tropical | 478 | $1.20 \times 10^{-2}$ |
| 9 | Water Body | 10 | $1.57 \times 10^{-2}$ |

During monsoon (June, July and August) 2013, majority of the lightning occurred in between the altitude of range (70-2000) m which contains the ecological zones namely Lower Tropical, Sub-tropical and Upper Tropical. These three zones respectively had 447, 467 and 478 numbers of lightning strikes. 478 of Upper Tropical zone is the highest value among nine zones where as a single value that belongs to Trans-Himalayan zone is the lowest value. Water Body zone whose number of strikes is 10, seems low in the column of number of lightning strikes has the second highest value of $1.57 \times 10^{-2}$/ km$^2$ in column with density of lightning strikes. The highest value of density of lightning strikes is $1.78 \times 10^{-2}$/ km$^2$ for Lower Tropical zone and the lowest value is calculated as $0.03 \times 10^{-2}$/ km$^2$ for Trans-Himalayan zone.

Figure 3.14 represents the lightning activity during pre-monsoon 2013 over different ecological zones of Nepal. In line graph, three peak points in the dotted line indicates dominance of the tropical zones namely Lower Tropical, Sub-tropical and Upper Tropical in number of lightning strikes..

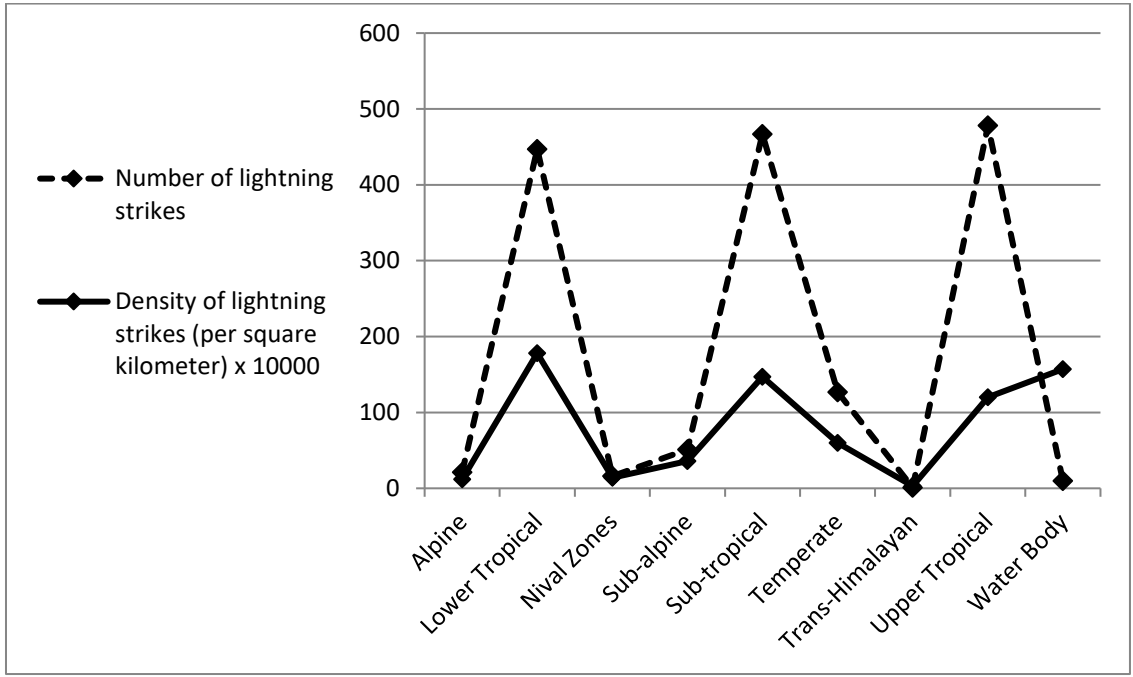

Figure 3.14: Lightning activity during monsoon 2013 over different ecological zones of Nepal.

**(D) Post-monsoon**

**(a) Post-monsoon 2012**

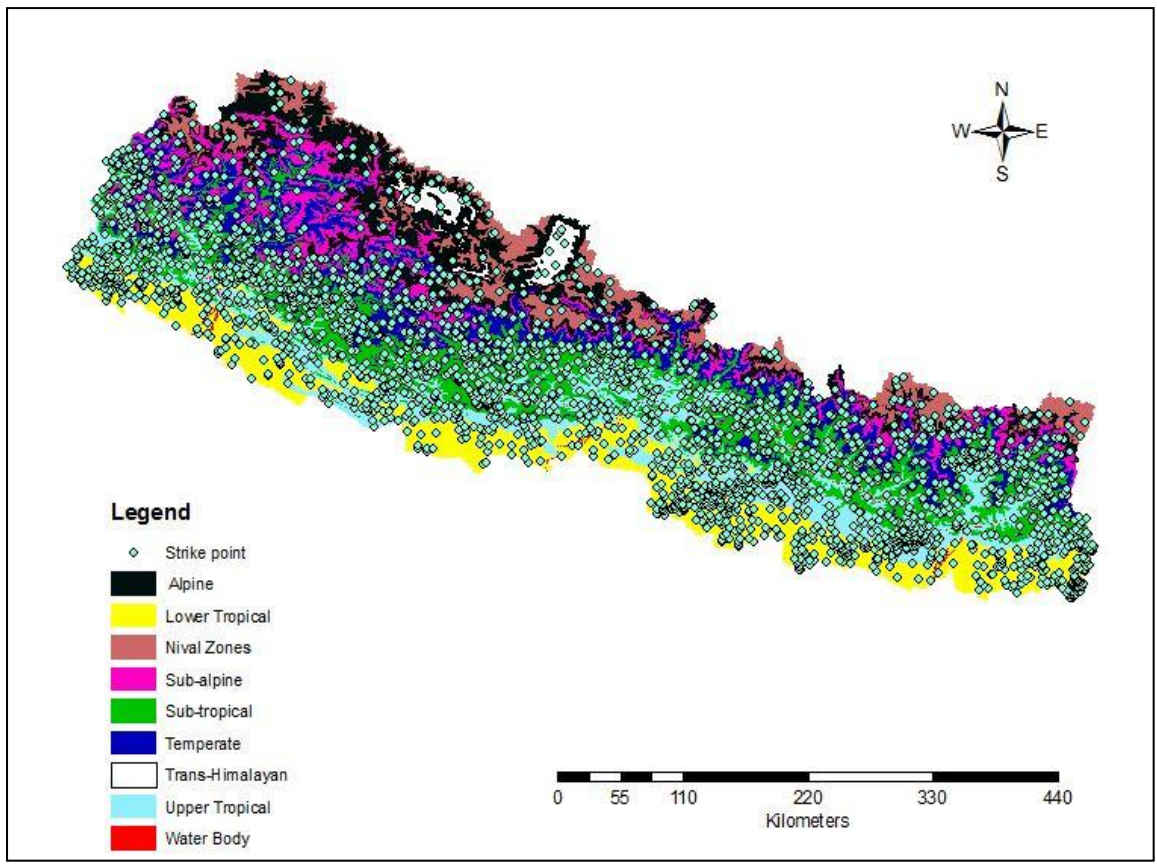

Figure 3.15: Ecological map of Nepal with lightning strikes during post-monsoon 2012.

Table 3.9: Lightning activity during post-monsoon 2012 over different ecological zones of Nepal.

| S.N | Ecological zones | Number of lightning strikes | Density of lightning strikes (per square kilometer) |
|---|---|---|---|
| 1 | Alpine | 98 | $0.57 \times 10^{-2}$ |
| 2 | Lower Tropical | 636 | $2.52 \times 10^{-2}$ |
| 3 | Nival Zones | 64 | $0.58 \times 10^{-2}$ |
| 4 | Sub-alpine | 172 | $1.20 \times 10^{-2}$ |
| 5 | Sub-tropical | 747 | $2.36 \times 10^{-2}$ |
| 6 | Temperate | 346 | $1.65 \times 10^{-2}$ |
| 7 | Trans-Himalayan | 21 | $0.65 \times 10^{-2}$ |
| 8 | Upper Tropical | 578 | $2.41 \times 10^{-2}$ |
| 9 | Water Body | 9 | $1.41 \times 10^{-2}$ |

The observation of table 3.9 which was created using the map obtained from incorporating data into ArcGIS shows that 2671 number of lightning strikes were recorded during post-monsoon (September, October and November) 2012. Continuing the trends of the other seasons of 2012, it was one of the tropical zones which experienced highest number of strikes. This time it happened to be Sub-tropical zone with 747 lightning strikes which are about 28 percent of total strikes. Lower Tropical and Upper Tropical zones do not have much difference values of number of lightning strikes on comparing with Sub-tropical zone. They were found to have 636 and 578 strikes respectively. Water Body zone was the one receiving lowest number of strikes. The largest value of density of lightning was calculated for Sub-tropical zone and Alpine was found to have smallest value for density of lightning after calculation.

Figure 3.16 depicts the information about lightning activity over different ecological zones on Nepal during post-monsoon 2012.

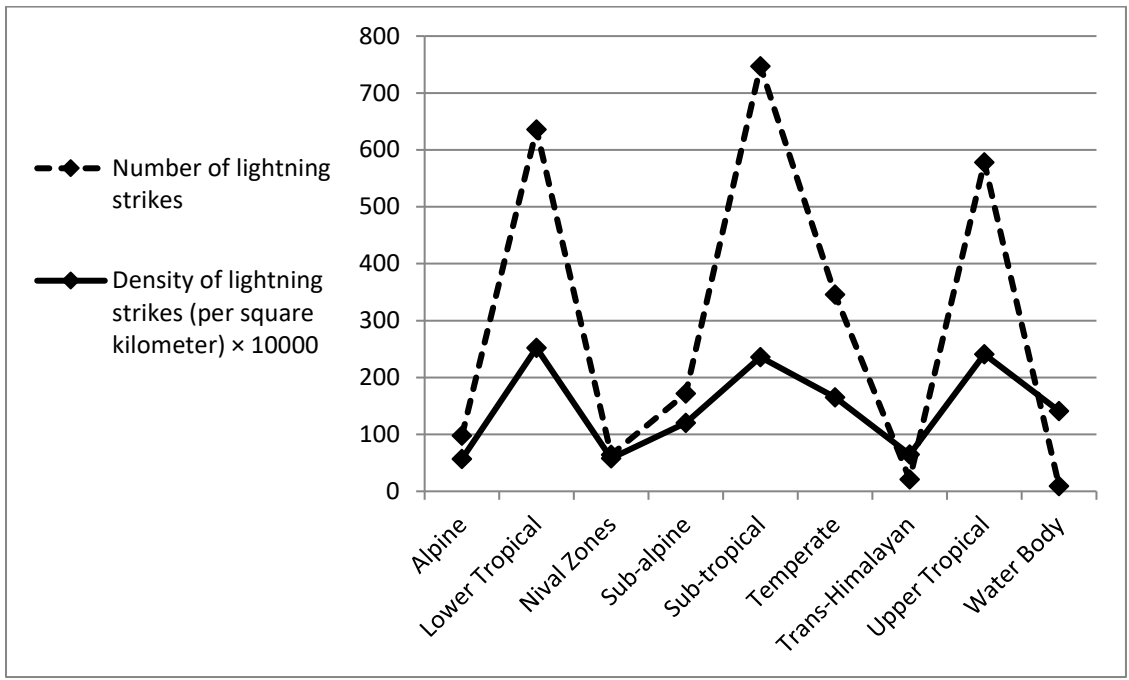

Figure 3.16: Lightning activity during post-monsoon 2012 over different ecological zones of Nepal.

**(b) Post-monsoon 2014 (October and November only)**

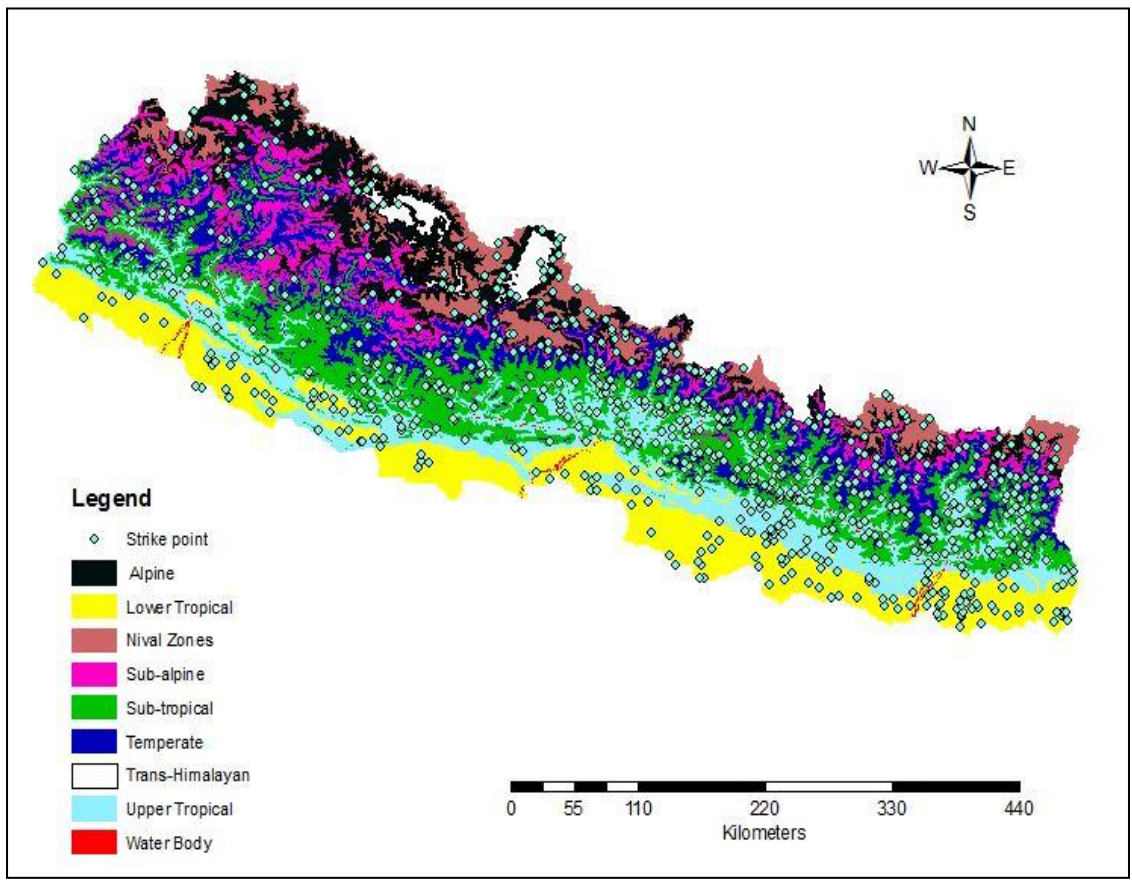

Figure 3.17: Ecological map of Nepal with lightning strikes for two months of post-monsoon (October, November) 2014.

Table 3.10: Lightning activity for two months of post-monsoon (October, November) 2014 over different ecological zones of Nepal.

| S.N | Ecological zones | Number of lightning strikes | Density of lightning strikes (per square kilometer) |
|---|---|---|---|
| 1 | Alpine | 75 | $0.44 \times 10^{-2}$ |
| 2 | Lower Tropical | 124 | $0.49 \times 10^{-2}$ |
| 3 | Nival Zones | 37 | $0.33 \times 10^{-2}$ |
| 4 | Sub-alpine | 69 | $0.48 \times 10^{-2}$ |
| 5 | Sub-tropical | 208 | $0.66 \times 10^{-2}$ |
| 6 | Temperate | 112 | $0.54 \times 10^{-2}$ |
| 7 | Trans-Himalayan | 11 | $0.34 \times 10^{-2}$ |
| 8 | Upper Tropical | 152 | $0.64 \times 10^{-2}$ |
| 9 | Water Body | 1 | $0.16 \times 10^{-2}$ |

We used the map obtained by incorporating the strike data of October and November 2014 into ArcGIS in order to observe the lightning activity during post-monsoon 2014. When the available data were processed through ArcGIS we found the similar scenario while checking the regions for highest number of strikes with that of post-monsoon 2012. In 2012, Sub-tropical zone recorded highest number of strikes followed by Lower Tropical zone and Upper Tropical zone where as post-monsoon values for 2014 shows that Sub-tropical zone remains in the top and Lower and Upper Tropical zones inter change their position in comparison to 2012. Nevertheless, these three remains at top three in receiving higher number of lightning strikes. Water Body received the least number of strikes for the time period under consideration. From the column of density of lightning strikes, we can report that tropical zones namely Lower Tropical, Sub-tropical and Upper Tropical have values in the higher values.

Figure 3.18 has been plotted to view and compare the lightning activity over different ecological zones of Nepal during two months of post-monsoon 2014.

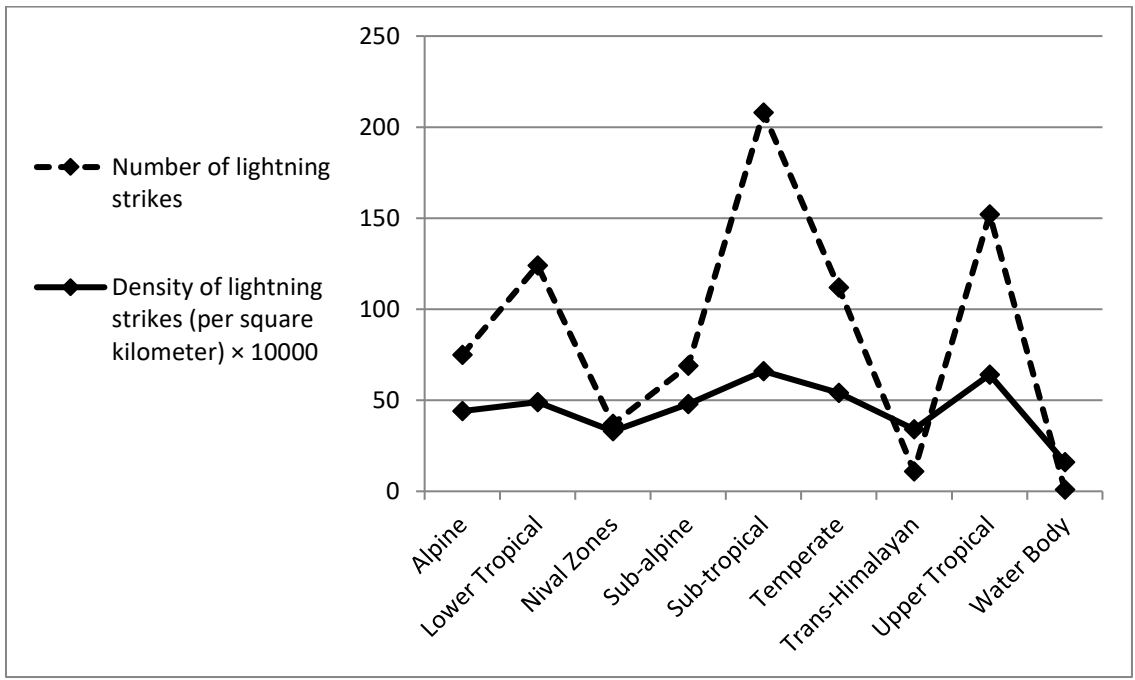

Figure 3.18: Lightning activity for two months of post-monsoon (October, November) 2014 over different ecological zones of Nepal.

# Chapter 4

# Discussion

Collection of lightning data, converting them to useable excel file for ArcGIS, making use of excel files to project and clip lightning data into Map of Nepal, spatial join of data to map and the observations from the finalized ecological map helped to calculate the results of lightning activity over different ecological zones for different seasons (excluding data from September 2013 to September 2014 as lightning sensor system was under repair.). In this chapter, we have calculated the average number of strikes and average density of lightning using arithmetic mean (AM). If the range of data varies by multiple of 10 or more, average is obtained using geometric mean otherwise arithmetic is preferred and under our study almost all range of data to calculate average do not vary with multiple of 10 or more. So, arithmetic mean has been used. Arithmetic mean is calculated by dividing the sum of the items by number of items. In the figures 4.1 to 4.9, listed in this chapter, unit for the density of lightning strikes is per square kilometre.

## 4.1 Alpine

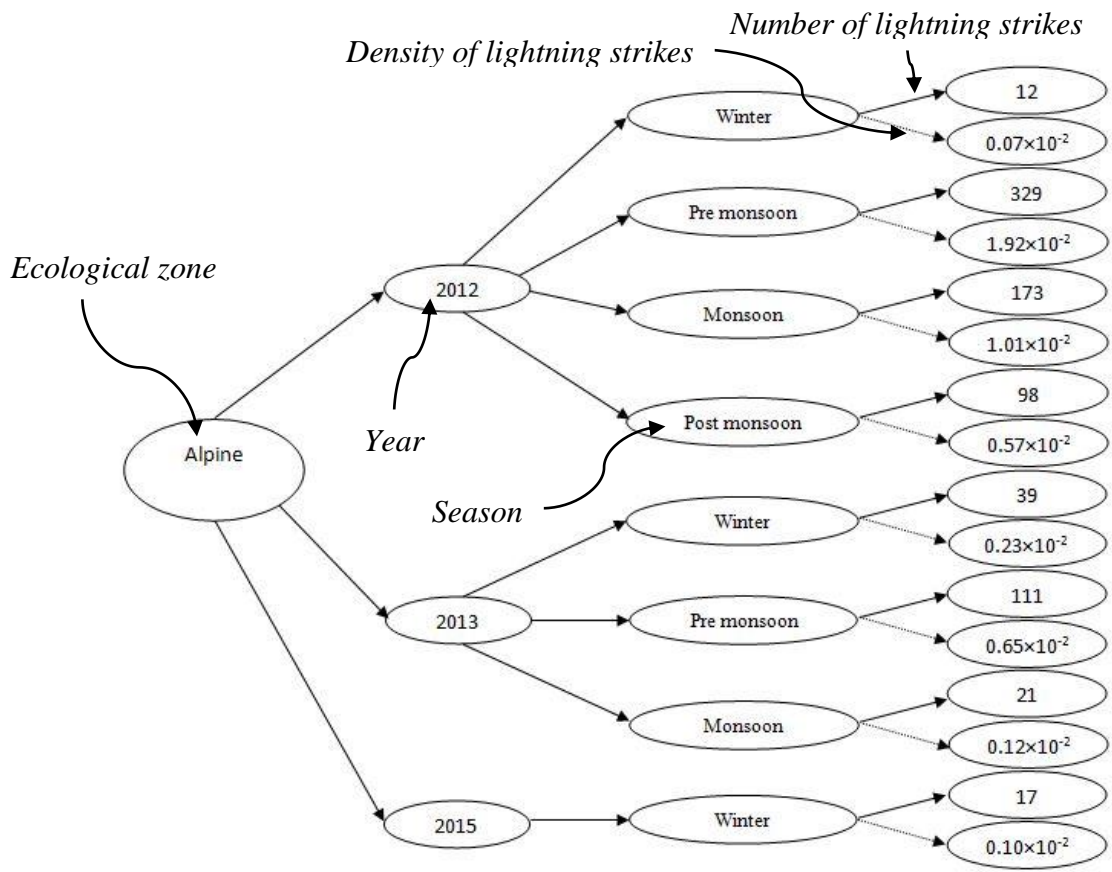

Figure 4.1: Schematic diagram of lightning activity over Alpine zone for different seasons for the years 2012, 2013 and 2015.

Number of lightning strikes and their densities for different seasons are obtained using formula for arithmetic mean.

For winter,

Average number of lightning strikes $= \dfrac{12 + 39 + 17}{3} = 22.67 \sim 23$ (approx)

Average density of lightning strikes $= \dfrac{(0.07 + 0.23 + 0.10) \times 10^{-2}}{3} = 0.13 \times 10^{-2}$ / km$^2$

Hence, approximately 23 lightning strikes occurred during winter with density of $0.13 \times 10^{-2}$ / km$^2$ over Alpine zone.

Now, for pre-monsoon,

Average number of lightning strikes $= \dfrac{329 + 111}{2} = 220$

Average density of lightning strikes $= \dfrac{(1.92 + 0.65) \times 10^{-2}}{2} = 1.29 \times 10^{-2}$ / km$^2$

Thus, during pre-monsoon, 220 strikes were recorded with density of $1.29 \times 10^{-2}$ / km$^2$ over Alpine zone in average.

Again, for monsoon,

Average number of lightning strikes $= \dfrac{173 + 21}{2} = 97$

Average density of lightning strikes $= \dfrac{(1.01 + 0.12) \times 10^{-2}}{2} = 0.57 \times 10^{-2}$/ km$^2$

As we see, 97 strikes of lightning with density of $0.57 \times 10^{-2}$ / km$^2$ can be expected over Alpine zone during monsoon.

Finally, for post-monsoon,

As we have single lightning activity of post-monsoon season of 2012, we can say that the average number and density of strikes during post-monsoon are found to be 98 and $0.57 \times 10^{-2}$/ km$^2$ respectively over Alpine zone.

By adding values of four different seasons, we can report that Alpine zone has trend of receiving around 438 lightning strikes per year with density of $2.56 \times 10^{-2}$ / km$^2$.

## 4.2 Lower Tropical

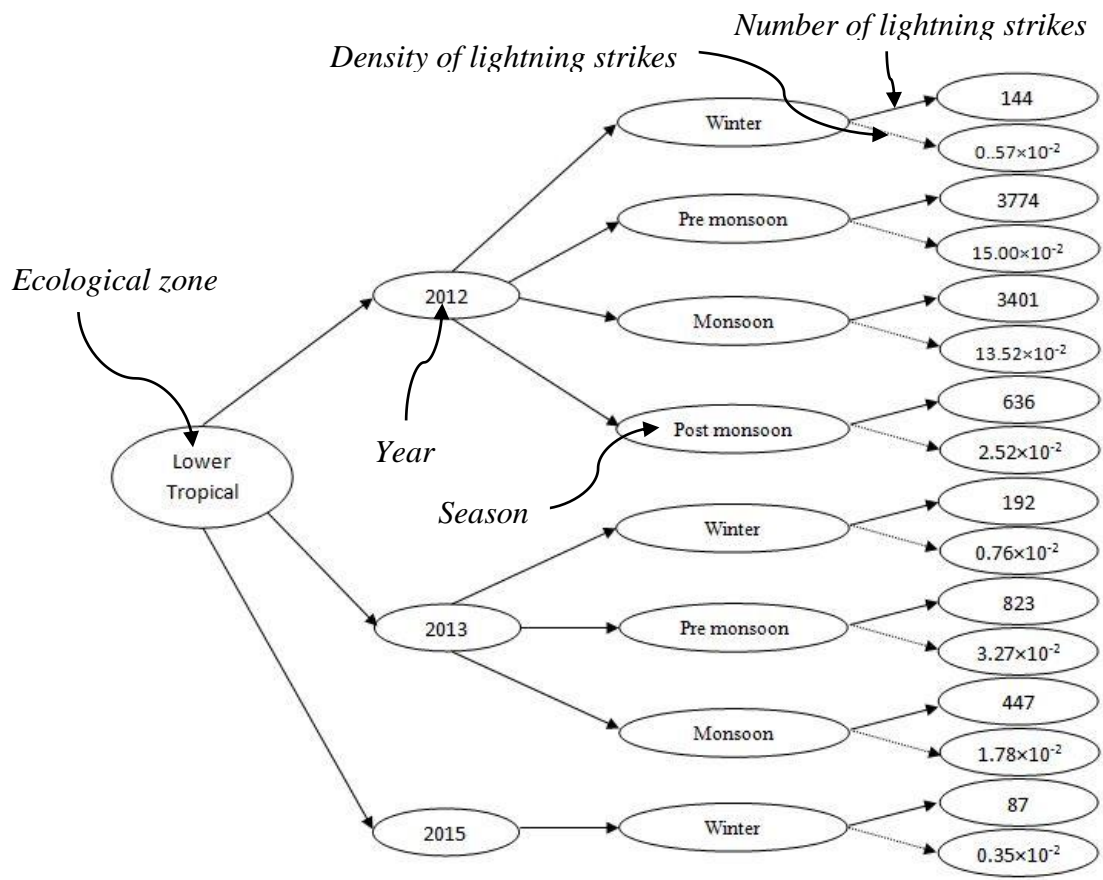

Figure 4.2: Schematic diagram of lightning activity over Lower Tropical zone for different seasons for the years 2012, 2013 and 2015.

Number of lightning strikes and their densities for different seasons are obtained using formula for arithmetic mean.

For winter,

Average number of lightning strikes = $\dfrac{144 + 192 + 87}{3} = 141$

Average density of lightning strikes = $\dfrac{(0.57 + 0.76 + 0.35) \times 10^{-2}}{3} = 0.56 \times 10^{-2} / \text{km}^2$

Hence, approximately 141 lightning strikes occurred during winter with density of $0.56 \times 10^{-2}/\text{km}^2$ over Lower Tropical zone.

Now, for pre-monsoon,

Average number of lightning strikes = $\dfrac{3774 + 823}{2} = 2299$

Average density of lightning strikes = $\dfrac{(15.00 + 3.27) \times 10^{-2}}{2} = 9.14 \times 10^{-2}/\text{km}^2$

Thus, during pre-monsoon, 2299 strikes were recorded with density of $9.14 \times 10^{-2}$/ km² over Lower Tropical zone in average.

Again, for monsoon,

Average number of lightning strikes = $\dfrac{3401 + 447}{2}$ =1924

Average density of lightning strikes = $\dfrac{(13.51 + 1.78) \times 10^{-2}}{2} = 7.65 \times 10^{-2}$ / km²

As we see, 1924 strikes of lightning with density of $7.65 \times 10^{-2}$ / km² can be expected over Lower Tropical zone during monsoon.

Finally, for post-monsoon,

As we have single lightning activity of post-monsoon season of 2012, we can say that the average number and density of strikes during post-monsoon are 636 and $2.52 \times 10^{-2}$/ km² respectively over Lower Tropical zone.

By adding values of four different seasons, we can report that Lower Tropical zone has trend of receiving around 3000 lightning strikes per year with density of $19.87 \times 10^{-2}$/ km².

## 4.3 Nival Zones

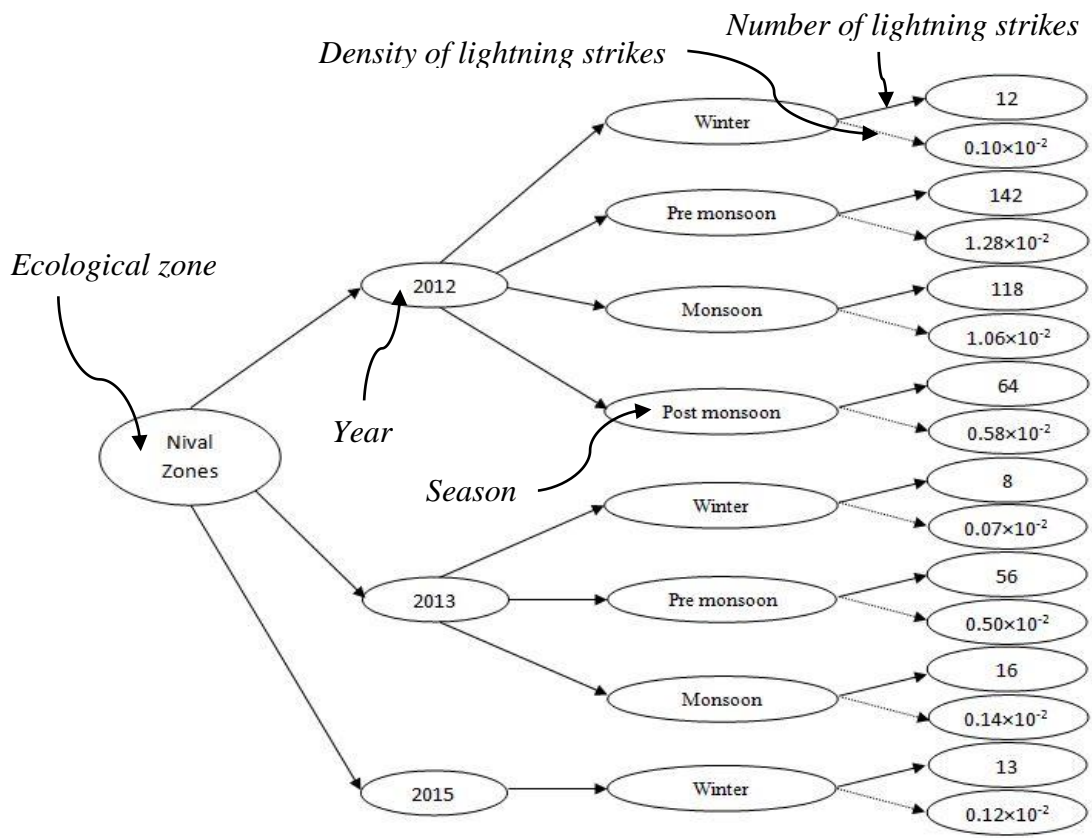

Figure 4.3: Schematic diagram of lightning activity over Nival Zones for different seasons for the years 2012, 2013 and 2015.

Number of lightning strikes and their densities for different seasons are obtained using formula for arithmetic mean.

For winter,

Average number of lightning strikes $= \dfrac{12 + 8 + 13}{3} = 11$

Average density of lightning strikes $= \dfrac{(0.10 + 0.07 + 0.12) \times 10^{-2}}{3} = 0.10 \times 10^{-2} / \text{km}^2$

Hence, approximately 11 lightning strikes occurred during winter with density of $0.10 \times 10^{-2}/ \text{km}^2$ over Nival Zones.

Now, for pre-monsoon,

Average number of lightning strikes $= \dfrac{142 + 56}{2} = 99$

Average density of lightning strikes $= \dfrac{(1.28 + 0.50) \times 10^{-2}}{2} = 0.89 \times 10^{-2}/ \text{km}^2$

Thus, in pre-monsoon 99 strikes were recorded with density of $0.89 \times 10^{-2}/ \text{km}^2$ over Nival Zones in average.

Again, for monsoon,

Average number of lightning strikes $= \dfrac{118 + 16}{2} = 67$

Average density of lightning strikes $= \dfrac{(1.06 + 0.14) \times 10^{-2}}{2} = 0.60 \times 10^{-2}/ \text{km}^2$

As we see, 67 strikes of lightning with density of $0.60 \times 10^{-2} / \text{km}^2$ can be expected over Nival Zones during monsoon.

Finally, for post-monsoon,

As we have single lightning activity of post-monsoon season of 2012, we can say that the average number and density of strikes during post-monsoon are 64 and $0.58 \times 10^{-2}/ \text{km}^2$ respectively over Nival Zones.

By adding values of four different seasons, we can report that Nival Zones has trend of receiving around 241 lightning strikes per year with density of $2.17 \times 10^{-2} / \text{km}^2$.

## 4.4 Sub-alpine

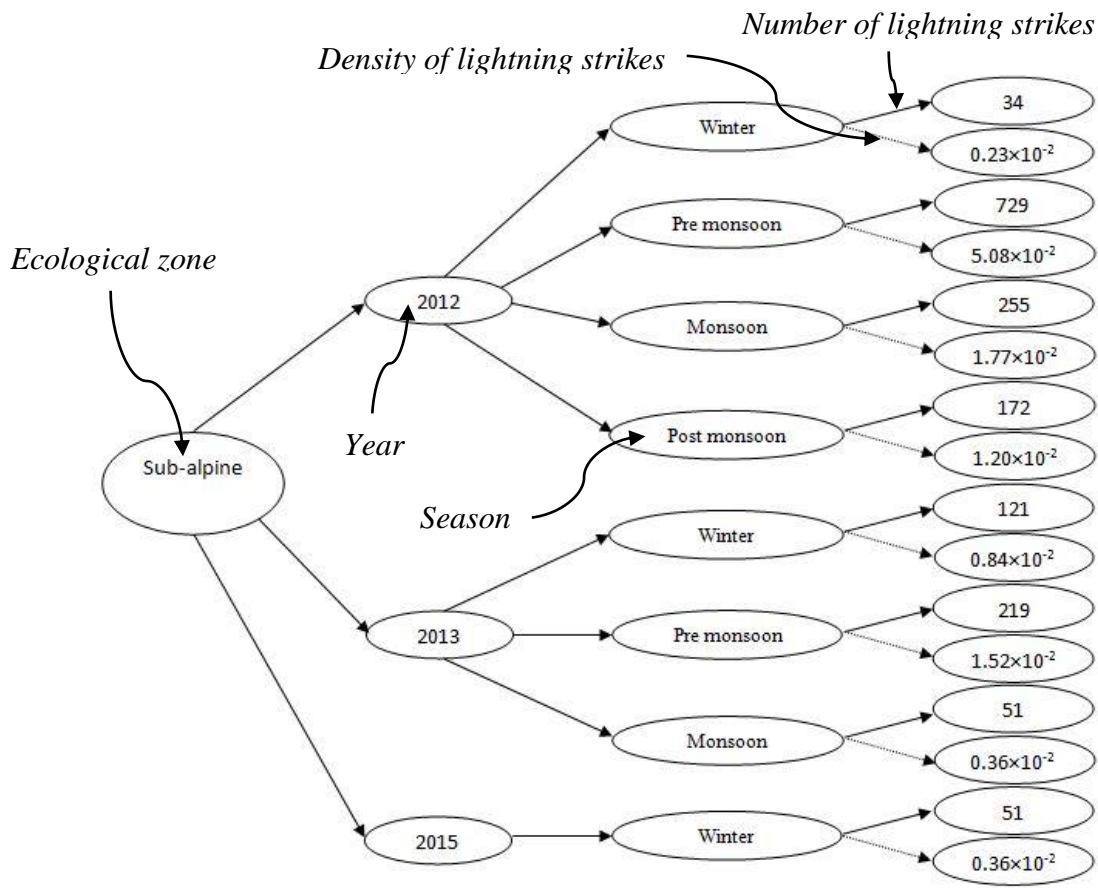

Figure 4.4: Schematic diagram of lightning activity over Sub-alpine zone for different seasons for the years 2012, 2013 and 2015.

Number of lightning strikes and their densities for different seasons are obtained using formula for arithmetic mean.

For winter,

Average number of lightning strikes $= \dfrac{34 + 121 + 51}{3} = 68.67 \sim 69$ (approx)

Average density of lightning strikes $= \dfrac{(0.23 + 0.84 + 0.36) \times 10^{-2}}{3} = 0.48 \times 10^{-2} / \text{km}^2$

Hence, approximately 69 lightning strikes occurred during winter with density of $0.48 \times 10^{-2} / \text{km}^2$ over Sub-alpine zone.

Now, for pre-monsoon,

Average number of lightning strikes $= \dfrac{729 + 219}{2} = 474$

Average density of lightning strikes $= \dfrac{(5.08 + 1.52) \times 10^{-2}}{2} = 3.30 \times 10^{-2} / \text{km}^2$

Thus, during pre-monsoon 474 strikes were recorded with density of $3.30 \times 10^{-2}/$ km² over Sub-alpine zone in average.

Again, for monsoon,

Average number of lightning strikes $= \dfrac{255 + 51}{2} = 153$

Average density of lightning strikes $= \dfrac{(1.77 + 0.36) \times 10^{-2}}{2} = 1.07 \times 10^{-2}/\,\text{km}^2$

As we see, 153 strikes of lightning with density of $1.07 \times 10^{-2}$ / km² can be expected over Sub-alpine zone during monsoon.

Finally, for post-monsoon,

As we have single lightning activity of post-monsoon season of 2012, we can say the average number and density of strikes in Sub-alpine zone during post-monsoon are 172 and $1.20 \times 10^{-2}/$ km² respectively.

By adding values of four different seasons, we can report that Sub-alpine zone has trend of receiving around 868 lightning strikes per year with density of $6.05 \times 10^{-2}/$ km².

## 4.5 Sub-tropical

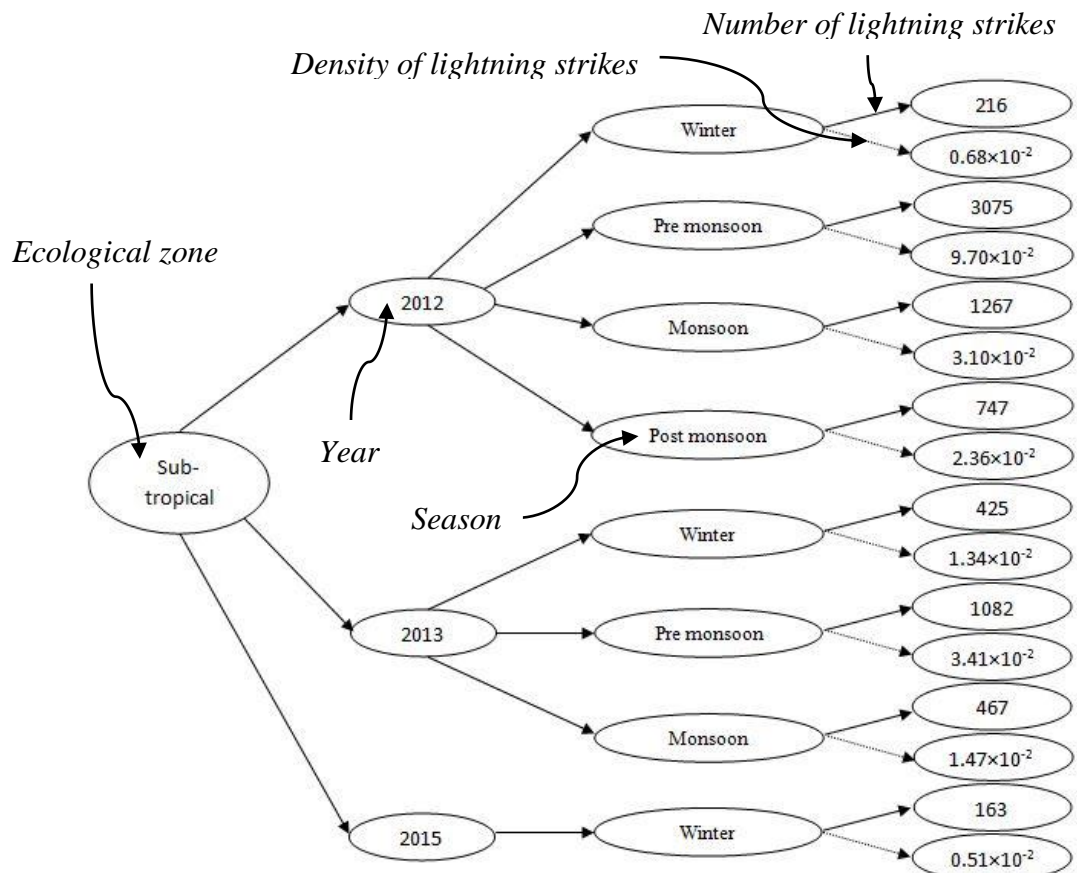

Figure 4.5: Schematic diagram of lightning activity over Sub-tropical zone for different seasons for the years 2012, 2013 and 2015.

Number of lightning strikes and their densities for different seasons are obtained using formula for arithmetic mean.

For winter,

Average number of lightning strikes $= \dfrac{216 + 425 + 163}{3} = 268$

Average density of lightning strikes $= \dfrac{(0.68 + 01.34 + 0.51) \times 10^{-2}}{3} = 0.84 \times 10^{-2} / \text{km}^2$

Hence, approximately 268 lightning strikes occurred during winter with density of $0.84 \times 10^{-2}/ \text{km}^2$ over Sub-tropical zone.

Now, for pre-monsoon,

Average number of lightning strikes $= \dfrac{3075 + 1082}{2} = 2079$

Average density of lightning strikes $= \dfrac{(9.70 + 3.41) \times 10^{-2}}{2} = 6.56 \times 10^{-2} / \text{km}^2$

Thus, during pre-monsoon, 2079 strikes were recorded with density of $6.56 \times 10^{-2}/ \text{km}^2$ over Sub-tropical zone in average.

Again, for monsoon,

Average number of lightning strikes $= \dfrac{1267 + 467}{2} = 867$

Average density of lightning strikes $= \dfrac{(3.10 + 1.47) \times 10^{-2}}{2} = 2.29 \times 10^{-2} / \text{km}^2$

As we see, 867 strikes of lightning with density of $2.29 \times 10^{-2} / \text{km}^2$ can be expected over Sub-tropical during monsoon.

Finally, for post-monsoon,

As we have single lightning activity of post-monsoon season of 2012, we can say that the average number and density of strikes over Sub-tropical zone during post-monsoon are 747 and $2.36 \times 10^{-2}/ \text{km}^2$ respectively.

By adding values of four different seasons, we can report that Sub-tropical zone has trend of receiving around 3961 lightning strikes per year with density of $12.05 \times 10^{-2} / \text{km}^2$.

## 4.6 Temperate

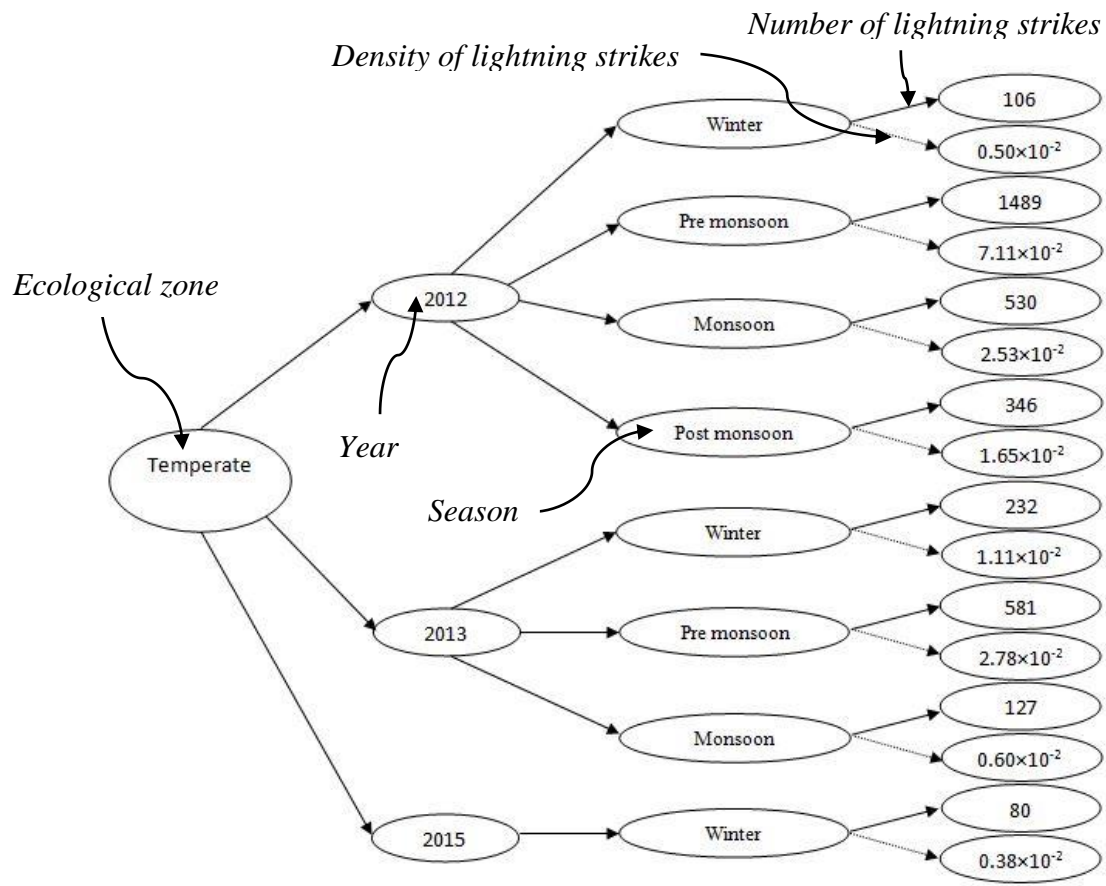

Figure 4.6: Schematic diagram of lightning activity over Temperate zone for different seasons for the years 2012, 2013 and 2015.

Number of lightning strikes and their densities for different seasons are obtained by using formula for arithmetic mean.

For winter,

Average number of lightning strikes $= \dfrac{106 + 232 + 80}{3} = 139.33 \sim 139$ (approx)

Average density of lightning strikes $= \dfrac{(0.50 + 1.11 + 0.38) \times 10^{-2}}{3} = 0.66 \times 10^{-2}$ / km$^2$

Hence, approximately 139 lightning strikes occurred during winter with density of $0.66 \times 10^{-2}$ / km$^2$ over Temperate zone.

Now, for pre-monsoon,

Average number of lightning strikes $= \dfrac{1489 + 581}{2} = 1035$

Average density of lightning strikes $= \dfrac{(7.11 + 2.78) \times 10^{-2}}{2} = 4.95 \times 10^{-2}$ / km$^2$

Thus, during pre-monsoon, 1035 strikes were recorded with density of $4.95 \times 10^{-2}$/ km² over Temperate zone in average.

Again, for monsoon,

Average number of lightning strikes $= \dfrac{530 + 127}{2} = 328.5 \sim 329$ (approx)

Average density of lightning strikes $= \dfrac{(2.53 + 0.60) \times 10^{-2}}{2} = 1.57 \times 10^{-2}$/ km²

As we see, 329 strikes of lightning with density of $1.57 \times 10^{-2}$/ km² can be expected over Temperate zone during monsoon.

Finally, for post-monsoon,

As we have single lightning activity of post-monsoon season of 2012, we can say that the average number and density of strikes over Temperate zone during post-monsoon are 346 and $1.65 \times 10^{-2}$ / km² respectively.

By adding values of four different seasons, we can report that Temperate zone has trend of receiving around 1849 lightning strikes per year with density of $8.83 \times 10^{-2}$ / km².

## 4.7 Trans-Himalayan

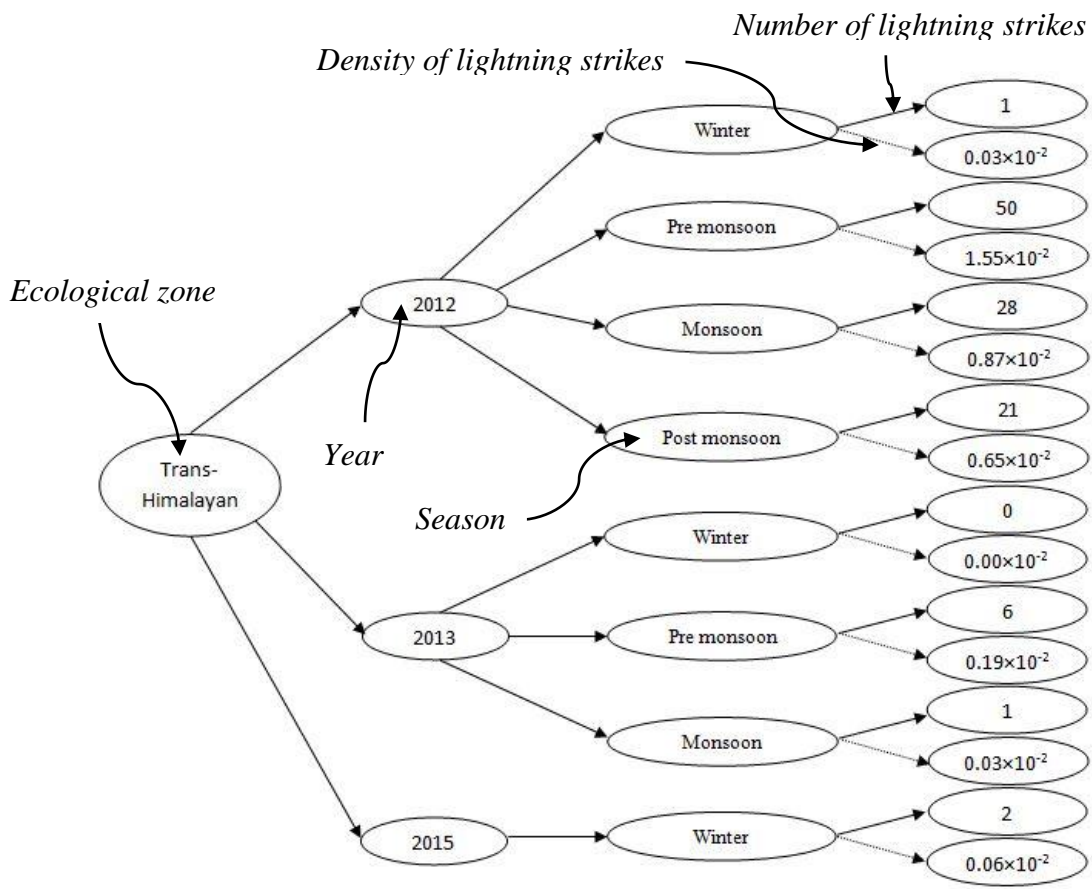

Figure 4.7: Schematic diagram of lightning activity over Trans-Himalayan for different seasons for the years 2012, 2013 and 2015.

Number of lightning strikes and their densities for different seasons are obtained using formula for arithmetic mean.

For winter,

Average number of lightning strikes $= \dfrac{1 + 0 + 2}{3} = 1$

Average density of lightning strikes $= \dfrac{(0.03 + 0.00 + 0.06) \times 10^{-2}}{3} = 0.03 \times 10^{-2}/\text{km}^2$

Hence, approximately only one lightning strike occurred during winter with density of $0.03 \times 10^{-2}/\text{km}^2$ over Trans-Himalayan zone.

Now, for pre-monsoon,

Average number of lightning strikes $= \dfrac{50 + 6}{2} = 28$

Average density of lightning strikes $= \dfrac{(1.55 + 0.19) \times 10^{-2}}{2} = 0.87 \times 10^{-2}/\text{km}^2$

Thus, during pre-monsoon, 28 strikes were recorded with density of $0.87 \times 10^{-2}/\text{km}^2$ over Trans-Himalayan zone in average.

Again, for monsoon,

Average number of lightning strikes $= \dfrac{28 + 1}{2} = 14.5$ ~15 (approx)

Average density of lightning strikes $= \dfrac{(0.87 + 0.03) \times 10^{-2}}{2} = 0.45 \times 10^{-2}/\text{km}^2$

As we see, 15 strikes of lightning with density of $0.45 \times 10^{-2}/\text{km}^2$ can be expected over Trans-Himalayan during monsoon.

Finally, for post-monsoon,

As we have single lightning activity of post-monsoon season of 2012, we can say that the average number and density of strikes during post-monsoon are 21 and $0.65 \times 10^{-2}/\text{km}^2$ respectively over Trans-Himalayan zone.

By adding values of four different seasons, we can report that Trans-Himalayan zone has trend of receiving around 65 lightning strikes per year with density of $2.00 \times 10^{-2}/\text{km}^2$.

## 4.8 Upper Tropical

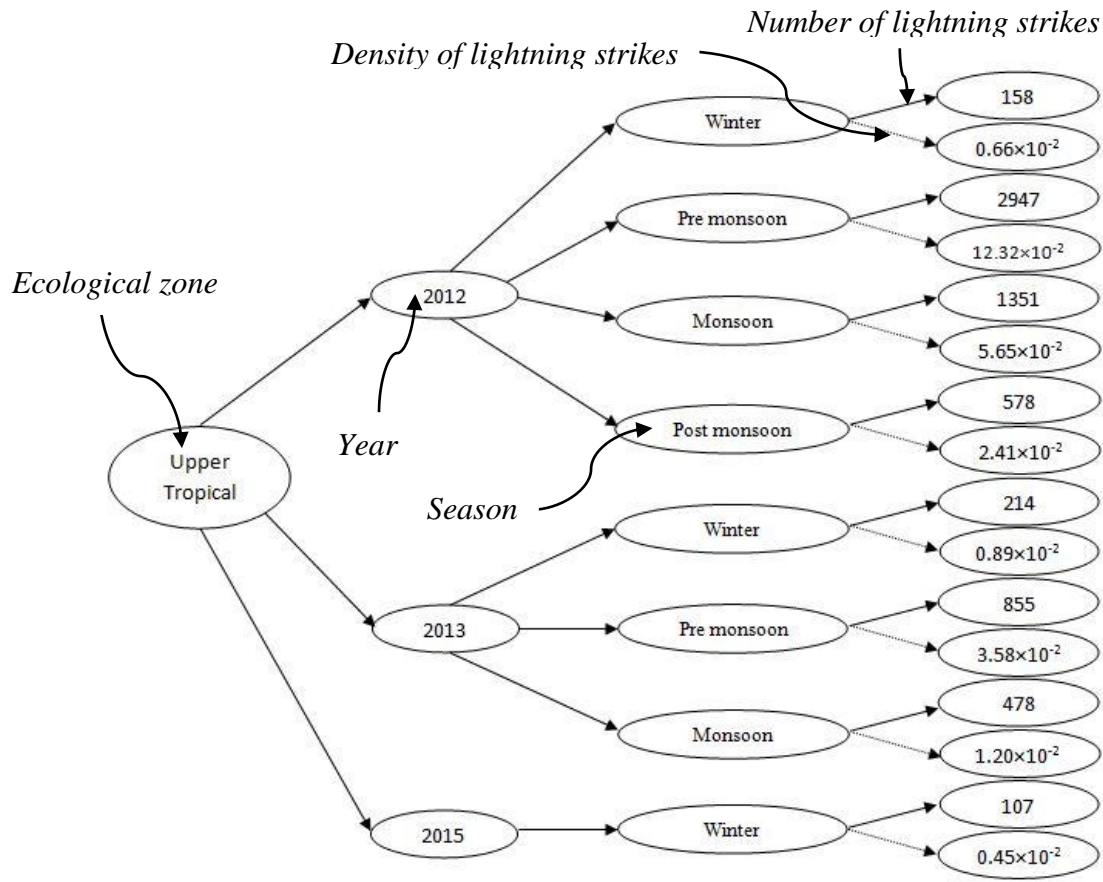

Figure 4.8: Schematic diagram of lightning activity over Upper Tropical zone for different seasons for the years 2012, 2013 and 2015.

Number of lightning strikes and their densities for different seasons are obtained using formula for arithmetic mean.

For winter,

Average number of lightning strikes $= \dfrac{158 + 214 + 107}{3} = 159.67 \sim 160$ (approx)

Average density of lightning strikes $= \dfrac{(0.66 + 0.89 + 0.45) \times 10^{-2}}{3} = 0.67 \times 10^{-2}$ / km$^2$

Hence, approximately 160 lightning strikes occurred during winter with density of $0.67 \times 10^{-2}$/ km$^2$ over Upper Tropical zone.

Now, for pre-monsoon,

Average number of lightning strikes $= \dfrac{2947 + 855}{2} = 1901$

Average density of lightning strikes $= \dfrac{(12.32 + 3.58) \times 10^{-2}}{2} = 7.95 \times 10^{-2}$/ km$^2$

Thus, during pre-monsoon, 1901 strikes were recorded with density of $7.95 \times 10^{-2}$/ km² over Upper Tropical zone in average.

Again, for monsoon,

Average number of lightning strikes $= \dfrac{1351 + 478}{2} = 914.5 \sim 915$ (approx)

Average density of lightning strikes $= \dfrac{(5.65 + 1.20) \times 10^{-2}}{2} = 3.43 \times 10^{-2}$ / km²

As we see, 915 strikes of lightning with density of $3.43 \times 10^{-2}$ / km² can be expected over Upper Tropical zone during monsoon.

Finally, for post-monsoon,

As we have single lightning activity of post-monsoon season of 2012, we can say that the average number and density of strikes during post-monsoon over Upper Tropical are 578 and $2.41 \times 10^{-2}$/ km² respectively.

By adding values of four different seasons, we can report that Upper Tropical zone has trend of receiving around 3554 lightning strikes per year with density of $14.46 \times 10^{-2}$/ km².

## 4.9 Water Body

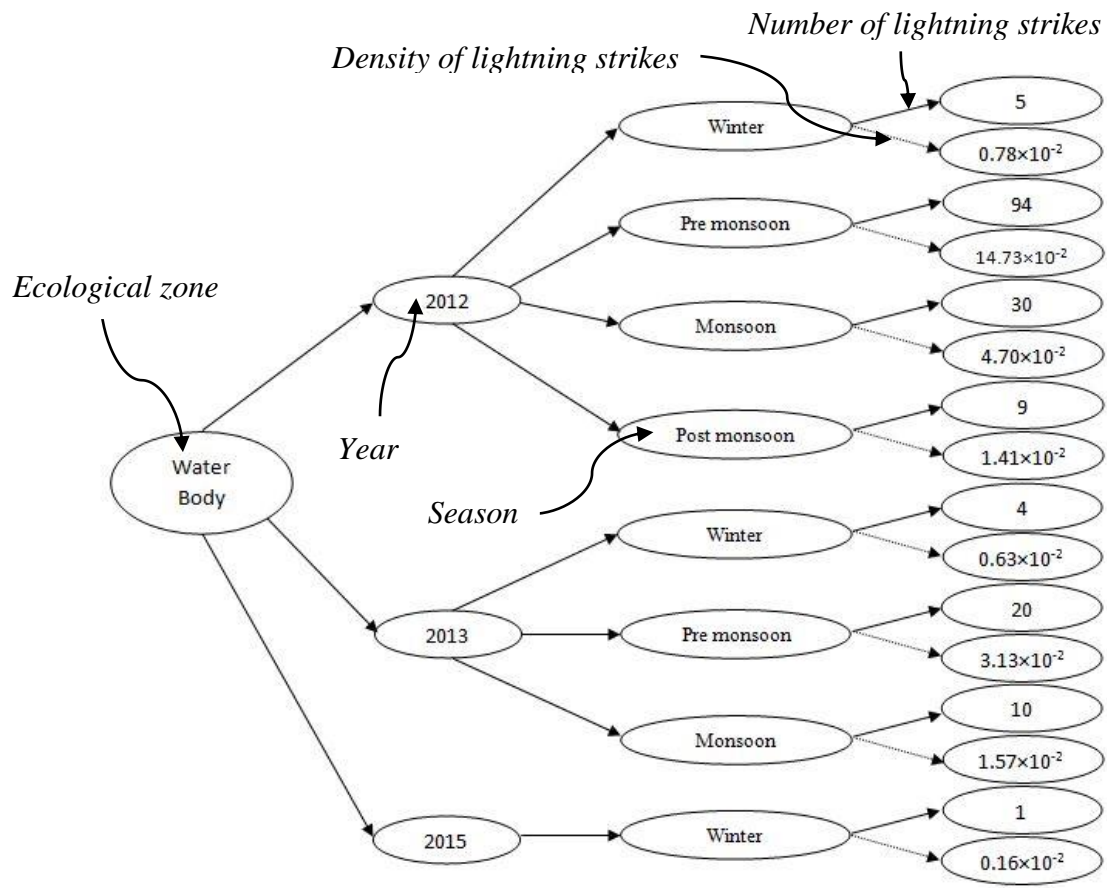

Figure 4.9: Schematic diagram of lightning activity over Water Body zone for different seasons for the years 2012, 2013 and 2015.

Number of lightning strikes and their densities for different seasons are obtained by using formula for arithmetic mean.

For winter,

Average number of lightning strikes $= \dfrac{5 + 4 + 1}{3} = 3.33 \sim 3$ (approx)

Average density of lightning strikes $= \dfrac{(0.78 + 0.63 + 0.16) \times 10^{-2}}{3} = 0.52 \times 10^{-2} / \text{km}^2$

Hence, approximately 3 lightning strikes occurred during winter with density of $0.52 \times 10^{-2} / \text{km}^2$ over Water Body zone.

Now, for pre-monsoon,

Average number of lightning strikes $= \dfrac{94 + 20}{2} = 57$

Average density of lightning strikes $= \dfrac{(14.73 + 3.13) \times 10^{-2}}{2} = 8.93 \times 10^{-2} / \text{km}^2$

Thus, during pre-monsoon, 57 strikes were recorded with density of $8.93 \times 10^{-2} / \text{km}^2$ over Water Body zone in average.

Again, for monsoon,

Average number of lightning strikes $= \dfrac{30 + 10}{2} = 20$

Average density of lightning strikes $= \dfrac{(4.70 + 1.57) \times 10^{-2}}{2} = 3.14 \times 10^{-2} / \text{km}^2$

As we see, 20 strikes of lightning with density of $3.14 \times 10^{-2} / \text{km}^2$ can be expected over Water Body zone during monsoon.

Finally, for post-monsoon,

As we have single lightning activity of post-monsoon season of 2012, we can say that the average number and density of strikes during post-monsoon over Water Body zone are 9 and $1.41 \times 10^{-2} / \text{km}^2$ respectively.

By adding values of four different seasons, we can report that Water Body zone has trend of receiving around 89 lightning strikes per year with density of $12.00 \times 10^{-2} / \text{km}^2$.

Table 4.1: Annually expected total number and density of lightning strikes over different ecological zones of Nepal.

| S.N | Ecological Zones | Number of Lightning strikes | Density of lightning strikes (per square kilometer) |
|-----|------------------|-----------------------------|-----------------------------------------------------|
| 1 | Alpine | 438 | $2.56 \times 10^{-2}$ |
| 2 | Lower Tropical | 3000 | $19.87 \times 10^{-2}$ |
| 3 | Nival Zones | 241 | $2.17 \times 10^{-2}$ |
| 4 | Sub-alpine | 868 | $6.05 \times 10^{-2}$ |
| 5 | Sub-tropical | 3961 | $12.05 \times 10^{-2}$ |
| 6 | Temperate | 1849 | $8.83 \times 10^{-2}$ |
| 7 | Trans-Himalayan | 65 | $2.00 \times 10^{-2}$ |
| 8 | Upper Tropical | 3554 | $14.46 \times 10^{-2}$ |
| 9 | Water Body | 89 | $12.00 \times 10^{-2}$ |

From the table 4.1, it can be summarised that Sub-tropical zone receives maximum number of lightning strikes with 3961 strikes per year but the annual highest density of lightning strikes is experienced by Lower Tropical zone with the value of $19.87 \times 10^{-2}$ strikes per square kilometer. Trans-Himalayan is the zone which has both the least values for number of strikes and densities of lightning strikes. It receives 65 lightning strikes with density of $2.00 \times 10^{-2}$ per square kilometer per year.

# Chapter 5

# Conclusion and future work

## 5.1 Conclusions

Based on our findings and the literature survey of similar topics, following conclusions can be drawn.

- **Lightning has direct relationship with surface temperature.**

    From the observations we had, we can completely agree with the narrated line "Recent studies continue to show the high positive correlation between surface temperatures and lightning activity" [14]. If we consider seven ecological zones of Nepal classified according to altitude, tropical zones viz. Lower Tropical zone, Sub-tropical zone and Upper Tropical zone, which ranges in between altitude of 70 to 2000 meters, are the zones with higher temperatures. While moving up towards higher altitude, temperature goes on decreasing for the ecological zones. For these seven zones, values of density of lightning strikes are in maximum range for the tropical zones and goes on decreasing for the higher altitudinal zones. In fact, Water Body zone, ecological zone which has not been classified on the basis of altitude, has comparable value of lightning density to Sub-tropical zone falls in the altitude range of 70 to 2000 meters. Thus, lightning is directly related with surface temperatures and can be represented with the graph below.

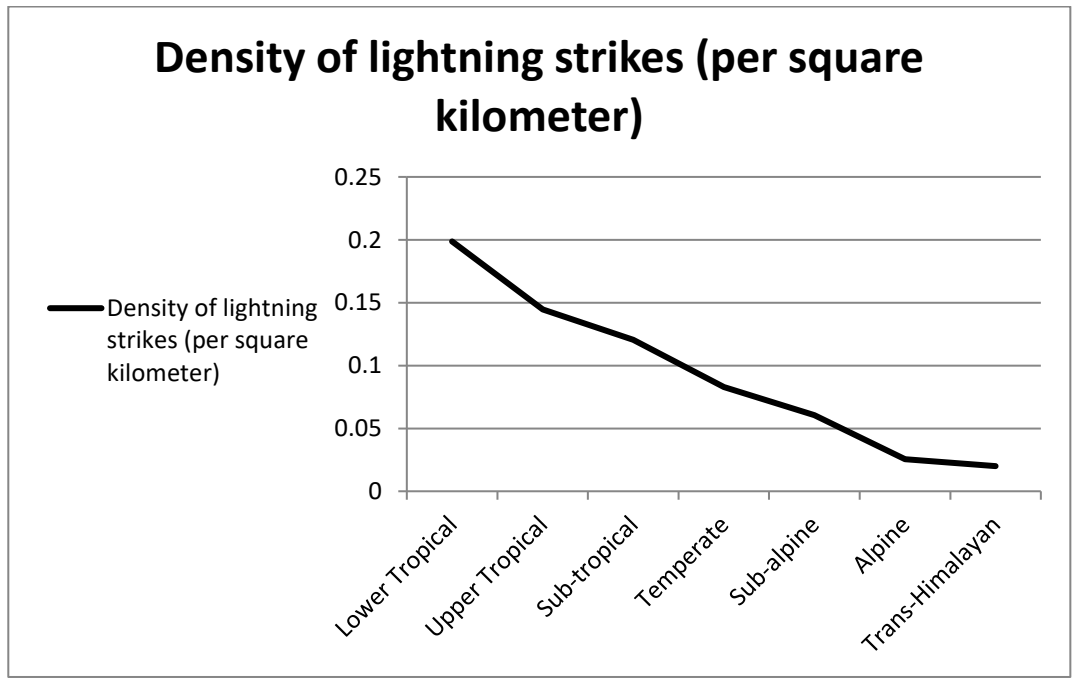

Figure 5.1: Density of lightning strikes over ecological zones in ascending altitude (temperature descending).

- **Lightning activities vary with seasons.**

    From the lightning activity observed over each ecological zone, it is easily noticeable that lightning activities go on changing throughout the year. As different seasons have different climatic conditions along with various amount and strength of storms, conclusion of different lightning in different season is easily predictable. This predictable conclusion has been approved from our observations. Generally, winter experiences the least lightning strikes, pre-monsoon records largest amount and then lightning strikes go on decreasing in monsoon and post-monsoon respectively. Hence, lightning activities change with seasons.

- **Presence of water vapour does play part to determine lightning activities over ecological zone.**

    It is seen that Water Body zone is hit by fewer lightning strikes among nine ecological zones while considering number of strikes but it has its value in higher side while seeing the densities of strikes. Among nine ecological zones, water vapour is present in largest amount in the surrounding atmosphere of Water Body and this water might have assisted to capture solar heat in order to increase surface temperature. Thus, we can conclude that amount of water vapour in any region has direct relationship with lightning activity.

- **Zones in the altitudinal region of 70 to 2000 meters need more lightning safety program to minimize effects.**

    Government and concerned authorities need to launch lightning awareness programme and install lightning protection system all over Nepal. But these programs are to be more frequent and protection systems are to be in installed at short intervals in Lower Tropical, Sub-tropical, Upper Tropical zones and Water Body zone. The basis of this conclusion is that these regions are the ones which bear higher density of lightning strikes and are also densely populated areas which can easily become vulnerable to disastrous effects of lightning.

**5.2 Recommendations for future study**

"**A study of lightning activity over different ecological zones of Nepal**" can be a research topic to explore more in future. Surface temperature has been concluded as the main factor contributing for the highest number and densities of strikes over the tropical zones namely Lower Tropical, Sub-tropical and Upper Tropical. Searching for other dominant causes that bring such a result can be an interesting matter to look for. Use of data of longer time period can minimize seen and unseen restrictions that might have occurred during the study. Hence, more data could be more than fruitful to obtain better result. Furthermore, we would like to recommend categorizing the available data of seasons into months and making analysis in monthly basis and then compare the results with previously performed research works done by categorizing data into seasons.